\documentclass[%
 reprint,
superscriptaddress,
nofootinbib,
 amsmath,amssymb,
 aps,
pra,
]{revtex4-1}
\usepackage{graphicx}
\usepackage{dcolumn}
\usepackage{bm}
\usepackage{hyperref}
\usepackage{amsmath}
\usepackage{cleveref}
\usepackage[inline]{enumitem}
\usepackage[font=small]{caption}
\usepackage[labelformat=empty, position=top]{subcaption}
\usepackage[export]{adjustbox}
\usepackage{blindtext}

\begin{document}


\title{Multiplexing rhythmic information by spike timing dependent plasticity}

\author{Nimrod Sherf}
  \email{sherfnim@post.bgu.ac.il}
  \affiliation{%
  	Physics Department, Ben-Gurion University of the Negev
  }%
\affiliation{%
	Zlotowski Center for Neuroscience, Ben-Gurion University of the Negev
}
\author{Maoz Shamir}%
 \email{shmaoz@bgu.ac.il}
\affiliation{%
Physics Department, Ben-Gurion University of the Negev
}%



\affiliation{%
Zlotowski Center for Neuroscience, Ben-Gurion University of the Negev
}

\affiliation{%
	Dept.\ of Physiology and Cell-biology, Faculty of Health Sciences, Ben-Gurion University of the Negev
}

\date{\today}

\begin{abstract}
Rhythmic activity has been associated with a wide range of cognitive processes including the encoding of sensory information, navigation, the transfer of emotional information and others. Previous studies have shown that spike-timing-dependent plasticity (STDP) can facilitate the transfer of rhythmic activity downstream the information processing pathway. However, STDP has also been known to generate strong winner-take-all like competitions between subgroups of correlated synaptic inputs. Consequently, one might expect that STDP would induce strong competition between different rhythmicity channels thus preventing the multiplexing of information across different frequency channels. This study explored whether STDP facilitates the multiplexing of information across multiple frequency channels, and if so, under what conditions. We investigated the STDP dynamics in the framework of a model consisting of two competing sub-populations of neurons that synapse in a feedforward manner onto a single post-synaptic neuron. Each sub-population was assumed to oscillate in an independent manner and in a different frequency band. To investigate the STDP dynamics, a mean field Fokker-Planck theory was developed in the limit of the slow learning rate. Surprisingly, our theory predicted limited interactions between the different sub-groups. Our analysis further revealed that the interaction between these channels was mainly mediated by the shared component of the mean activity. Next, we generalized these results beyond the simplistic model using numerical simulations. We found that for a wide range of parameters, the system converged to a solution in which the post-synaptic neuron responded to both rhythms. Nevertheless, all the synaptic weights remained dynamic and did not converge to a fixed point. These findings imply that STDP can support the multiplexing of rhythmic information, and demonstrate how functionality (multiplexing of information) can be retained in the face of continuous remodeling of all the synaptic weights. 

\begin{description}
\item[Key words]
 Hebbian learning; synaptic updating; WTA competition; multiplexing:
\end{description}
\end{abstract}

\pacs{Valid PACS appear here}
\maketitle


\section{\label{sec:Introduction}Introduction}

Neuronal oscillations have been described and studied for more than a century  \cite{coenen2014adolf,Haas2003,steriade1985abolition,buzsaki1992high,gray1994synchronous,bragin1999high,burgess2002functional,buzsaki2006rhythms,shamir2009representation,buzsaki2015editorial,ray2015gamma,bocchio2017synaptic,Proskovec2018,taub2018oscillations}. Rhythmic activity in the central nervous system has been associated with: attention, learning, encoding of external stimuli, consolidation of memory and motor output \cite{engel1992temporal,singer1995visual,engel2001dynamic,fries2005mechanism,buzsaki2006rhythms,jensen2007human,knyazev2007motivation,bocchio2017synaptic,hobson2002cognitive,shamir2009representation,engel2010beta,buzsaki2015editorial,taub2018oscillations,storchi2017modulation}. 
For example, Bocchio et al.\ \cite{bocchio2017synaptic} described how situations of perceived threat or perceived safety are translated into rhythmic activity in different frequency bands in different brain regions. The transfer of oscillatory signal between large neuronal populations is not trivial and requires some mechanism to prevent destructive interference and maintain the rhythmic component.

Recently, it was suggested that synaptic plasticity, and especially spike-timing-dependent plasticity (STDP), can provide such a mechanism. STDP can be thought of as a generalization of Hebb’s rule that neurons that “fire together wire together”  \cite{Hebb} to the temporal domain. In STDP, the amount of potentiation (increase in the synaptic strength) and depression (decrease in the synaptic strength) depends on the temporal relation between the pre- and post-synaptic firing\footnote{
	Neuronal interaction is not symmetric. A synaptic terminal connects two neurons: a transmitting neuron on the pre-synaptic side and a receiving neuron on the post-synaptic side. When a spike that propagates along the axon of the pre-synaptic neuron reaches the synaptic terminal it induces a change in the membrane potential of the post-synaptic neuron.}. Luz and Shamir \cite{luz2016oscillations} demonstrated that STDP can facilitate the transfer of rhythmic activity downstream, and analyzed  the features of the STDP that govern this ability.

In their study, Luz and Shamir focused on the transfer of rhythmic activity from a single population, thus ignoring the possibility of converging information from different brain regions. However, it has been suggested that rhythmic activity in different frequency bands is used for multiplexing information \cite{CFCExample1,CFCExample2,CFCExample3,CFCExample4,saleem2017subcortical}. On the other hand, STDP has been shown to generate a winner-take-all like competition between subgroups of correlated pools of neurons \cite{gilson2009emergence,gutig2003learning,morrison2008phenomenological,song2000competitive,shamir2019theories}. This raises the question whether STDP can enable the transmission of rhythmic activity in different frequency bands and facilitate multiplexing of information.

We address this question here in a modelling study. In  \cref{subsec:Architecture,subsec:neuroDyn,subsec:correl,subsec:STDP rule}, we define the network model and the STDP learning rule. Next, in \cref{subsec:analytical}, we derive a mean-field approximation for the STDP dynamics in the limit of a slow learning rate. Analysis of the STDP dynamics revealed constraints on the transmission of multi-frequency oscillations. In \cref{subsec:Numerical}, we extend our results using numerical simulations beyond the analytical toy model. Finally, in \cref{sec:Discussion}, we summarize our results and discuss their implications.

\section{\label{sec:results}Results}

\subsection{The pre-synaptic populations}  \label{subsec:Architecture}

\begin{figure}
	\centering
	\includegraphics[width=1\linewidth]{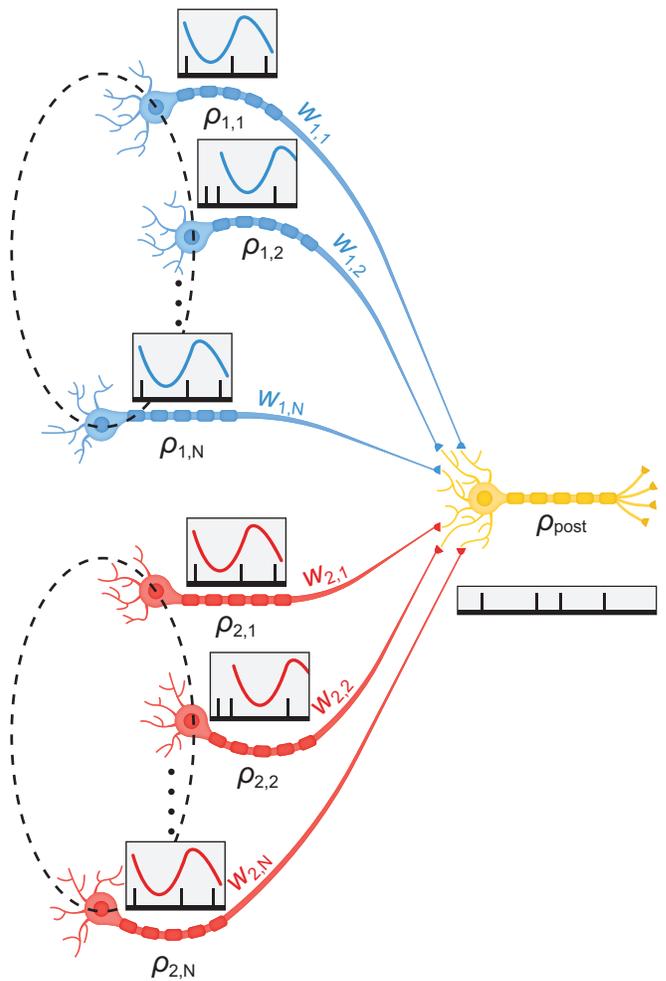}
	\caption{A schematic description of the network architecture showing two pre-synaptic populations, each oscillating at a different frequency. The output of these pre-synaptic neurons, serves as a feed-forward input to a single post-synaptic neuron.}
	\label{fig:twopopfignew}
\end{figure}

We model a system of two excitatory populations of $N$ neurons each, synapsing in a feed-forward manner onto the same downstream neuron.
 Both  populations are assumed to exhibit rhythmic activity in a different frequency, representing a different feature of the external stimulus. 
Each neuron is further characterized by a preferred phase, to which it is more likely to fire during the cycle. The preferred phases are assumed to be evenly spaced on the ring. Thus, it is convenient to think of the neurons in each population as organized on a ring according to their preferred phases of firing,  \cref{fig:twopopfignew}.

The spiking activity of neuron $k \in \left\lbrace 1, ... N \right\rbrace $ in population $\eta \in \left\lbrace 1,2  \right\rbrace$, $ \rho_{\eta,k}(t)=\sum_{i} \delta(t-t_{k,i}^\eta)$  is a doubly stochastic process, where $\lbrace t_{k,i}^\eta \rbrace _{i=1}^{\infty}$ are the spike times. Given the ‘intensity’ $D_\eta$ of feature  $\eta \in \left\lbrace 1,2 \right\rbrace$ of the stimulus, $ \rho_{\eta,k}(t)$, follows an independent inhomogeneous Poisson process statistics with a mean rate that is given by:
\begin{equation}\label{eq:meanfiring}
\langle \rho_{\eta,k}(t) \rangle=D_{\eta}(1+\gamma \cos[ \nu_{\eta}t-\phi_{\eta,k}]), \ \phi_{\eta,k}=2 \pi k/N.
\end{equation} 
where  $\nu_{\eta}$ is the angular frequency of oscillations for neurons in population $\eta$, $\gamma$ is the modulation to the mean ratio of the firing rate, and  $\phi_{\eta,k}$ is the phase of the $k$th neuron from population $\eta$.
As the intensity parameters, $D_\eta$, represent features of the external stimulus they also fluctuate on a timescale which is typically longer than the characteristic timescale of the neural response. For simplicity we assumed that $D_1$ and $D_2$ are independent random variables with an identical distribution:  

\begin{subequations} \label{eq:meanFR}
	\begin{align}
	&\langle D_{\eta}  \rangle =D, \\
	&\langle D_{\eta} D_{\xi}  \rangle =D^2(1+\sigma^2 \delta_{\eta \xi}),
	\end{align}
\end{subequations}
where $\langle ... \rangle$ denotes averaging with respect to the neuronal noise and stimulus statistics.
The essence of multiplexing is to enable the transmission of different information channels; hence, the assumption of independence of $D_1$ and $D_2$ represents fluctuations of different stimulus features. This assumption also drives the winner-take-all competition between the two populations.

\subsection{Post-synaptic neuron model}  \label{subsec:neuroDyn}

Spike time correlations are the driving force of STDP dynamics \cite{luz2014effect}. Correlated pairs are more likely to affect the  firing of the post-synaptic neuron, and as a result, to modify their synaptic connection \cite{gutig2003learning,shamir2019theories}.
To analyze the STDP dynamics we need a simplified model for the post-synaptic firing that will enable us to compute the pre-post cross-correlations, and in particular, their dependence on the synaptic weights.

 Following the work of \cite{luz2016oscillations,morrison2008,song2000competitive,kempter2001intrinsic,kempter1999hebbian,luz2012balancing}, the post-synaptic neuron is modeled as a linear Poisson neuron  with a characteristic delay $d>0$. The mean firing rate of the post-synaptic neuron at time $t$, $r_{post} (t)$,is given by
\begin{equation}\label{eq:meanpost}
r_{post} (t)=\frac{1}{N}\sum_{\eta=1}^{2} \sum_{k=1}^{N}w_{\eta,k}  \rho_{\eta,k}(t-d),
\end{equation}
where $w_{\eta,k}$ is the synaptic weight of the $k$th  neuron of population $\eta$.

\subsection{Temporal correlations \& order parameters} \label{subsec:correl}

The utility of the linear neuron model is that the pre-post correlations are given as a linear combination of the correlations of the pre-synaptic populations. The cross-correlation between pre-synaptic neurons is given by:
\begin{equation}\label{eq:correprepre}
\begin{split}
\Gamma_{(\eta,j), (\xi,k)}(\Delta t)&=\langle  \rho_{\eta,j}(t) \rho_{\xi,k}(t+\Delta t)   \rangle 
\\&= \delta_{\eta \xi}\bigg( D^2(1+\sigma^2)  \big(1  + 
\frac{\gamma^2}{2} \cos[ \nu_\eta \Delta t\\&
+\phi_{\eta,j}-\phi_{\xi,k}]\big)+ \delta_{j k} D \delta(\Delta t)\bigg) \\ & 
+D^2 (1-\delta_{\eta \xi}).
\end{split}
\end{equation}

The correlation between the $j$th neuron from population 1 and the post-synaptic neuron can therefore be written as

\begin{equation}\label{eq:corrFull}
\begin{split}
\Gamma_{(1,j), \text{ post}}(\Delta t)=&\frac{D}{N}\delta(\Delta t-d) w_{1,j}+
\frac{D^2}{N}\sum_{k=1}^{N}w_{1,k} 
 (1 +  \\ &   \sigma^2) \bigg(1  + 
\frac{\gamma^2}{2} \cos[ \nu_1 (\Delta t-d)+\phi_{1,j}- \\ & \phi_{1,k} ]\bigg) +\frac{D^2}{N}\sum_{l=1}^{N}w_{2,l}.
 \end{split}
\end{equation}
Thus, the correlations are determined by the global order parameters  $\bar{w}$ and $\tilde{w}e^{i \psi}$, where  $\bar{w}$ is  the mean synaptic weight and  $\tilde{w}e^{i \psi}$ is  its first Fourier component.  For $ N \gg 1$ these parameters are defined as follows
\begin{equation}\label{eq:wbar}
\bar{w}_\eta(t)= \int_{0}^{2 \pi}w_\eta(\phi, t)   \frac{d \phi}{2 \pi}
\end{equation}
and

\begin{equation}\label{eq:wtilda}
\tilde{w}_\eta(t) e^{i \psi_\eta}= \int_{0}^{2 \pi}w_\eta(\phi, t)e^{i\phi}\frac{d \phi}{2 \pi}.
\end{equation}
The phase $\psi_\eta$  is determined by the condition that $\tilde{w}_\eta$ is real non-negative.
Note that the coupling between the two populations is only expressed through the last term of the correlation function, \cref{eq:corrFull}.

\subsection{The STDP rule}  \label{subsec:STDP rule}

Based on  \cite{luz2014effect,gutig2003learning,luz2012balancing}  we model the synaptic modification $\Delta w$ following either a pre- or post-synaptic spike as:
\begin{equation}\label{eq:deltaw}
 \Delta{w}=\lambda[f_{+}(w)K_{+}(\Delta t)-f_{-}(w)K_{-}(\Delta t)],
\end{equation}
The STDP rule, \cref{eq:deltaw}, is written as a sum of two processes: potentiation (+, increase in the synaptic weight) and depression (-, decrease). We further assume a separation of variables by writing each process as a product of a weight dependent function, $f_\pm(w)$, and a temporal kernel, $K_\pm(\Delta t)$. The term $\Delta t = t_{post}-t_{pre}$ is the time difference between pre- and post-synaptic spiking. Here we assumed, for simplicity, that all pairs of pre and post spike times contribute additively to the learning process via \cref{eq:deltaw}. Note, however, that the temporal kernels of the STDP rule, $K_{\pm}(\Delta t)$ have a finite support. Here we normalized the kernels, $\int K_{\pm}(\Delta t) d\Delta t = 1$. The parameter $\lambda$ is the learning rate. It is assumed that the learning process is slower than the neuronal spiking activity and the timescale of changes in the external stimulus.
Here, we used the synaptic weight dependent functions of the form of \cite{gutig2003learning}: 
\begin{align}\label{eq:fplusminus}
f_{+}(w)&=(1-w)^{\mu} \\
f_{-}(w)&=\alpha w^{\mu}, 
\end{align}
where $\alpha>0$ is the relative strength of depression and $\mu\in[0,1]$ controls the non-linearity of the learning rule; thus, $f(w)_\pm$ ensures that the synaptic weights are confined to the boundaries $w \in [0,1]$. G\"{u}tig and colleagues \cite{gutig2003learning} showed that the relevant parameter regime for the emergence of a non-trivial structure is $\alpha>1$ and small $\mu$. 

Empirical studies reported a large repertoire of temporal kernels for STDP rules \cite{markram1997regulation,bi1998synaptic,sjostrom2001rate,zhang1998critical,Abbott2000,froemke2006contribution,Nishiyama2000CalciumSR,shouval2002unified,WOODIN2003807}. Here we focus on two families of STDP rules:
\begin{enumerate*}
	\item  A temporally asymmetric kernel  \cite{bi1998synaptic,zhang1998critical,Abbott2000,froemke2006contribution}.
	\item A temporally symmetric kernel \cite{Nishiyama2000CalciumSR,Abbott2000,shouval2002unified,WOODIN2003807}.
\end{enumerate*}

For the temporally asymmetric kernel we use the exponential model: 
\begin{equation}\label{eq:kernel}
K_{\pm}(\Delta t)=\frac{e^{\mp  \Delta t/\tau_{\pm}}}{\tau_{\pm}}\Theta(\pm  \Delta t),
\end{equation} 
where  $\Delta t=t_{post}-t_{pre}$, $\Theta(x)$ is the Heaviside function, and  $\tau_{\pm}$ is the characteristic timescale of the potentiation $(+)$ or depression $(-)$. We assume that $\tau_- > \tau_+$ as typically reported. 

For the temporally symmetric learning rule we use a difference of Gaussians model:
\begin{equation}\label{eq:kernelSymmetric}
K_{\pm}(\Delta t)=\frac{1}{\tau_\pm \sqrt{2 \pi}} e^{-\frac{1}{2} (\frac{\Delta t}{\tau_\pm})^2},
\end{equation} 
where $\tau_\pm$ is the temporal width. In this case, the order of firing is not important; only the absolute time difference.
We further assume, in both models, that $\tau_+ < \tau_-$, as is typically reported.

\subsection{STDP dynamics in the limit of slow learning}  \label{subsec:analytical}

Due to noisy neuronal activities, the learning dynamics is stochastic . However, in the limit of a slow learning rate, $\lambda \rightarrow 0$, the fluctuations become negligible and one can obtain deterministic dynamic equations for the (mean) synaptic weights (see \cite{luz2014effect} for a detailed derivation)
\begin{equation}\label{eq:wdot}
    \frac{\dot{w}_{\eta,j}(t)}{\lambda}=I^+_{(\eta,j)}(t) -I^-_{(\eta,j)}(t) 
\end{equation}
where $\eta=1,2$ and
\begin{equation}\label{eq:wdotpm}
    I^\pm_{(\eta,j)}(t)=f_{\pm}(w_{\eta,j}(t)) \int_{-\infty}^{\infty}\Gamma_{(\eta,j), \text{ post}}( \Delta)K_{\pm}(\Delta)d\Delta.
\end{equation}
%
%
Using the correlation structure, \cref{eq:corrFull},  \cref{eq:wdotpm} yields
\begin{equation}\label{eq:PotetiationDepression}
\begin{split}
    &I^\pm_{(\eta,j)}(t) \equiv    f_{\pm}(w_{\eta,j}(t)) D^2 \bar{K}_\pm\bigg( \bar{w}_{\eta}(t)(1+\sigma^2)+\bar{w}_{\xi} +\frac{\gamma^2}{2}\\ 
    &(1+\sigma^2) \frac{\tilde{K}_\pm}{\bar{K}_\pm} \tilde{w}_\eta(t) \cos[\phi_{\eta,j}-\Omega^\eta_\pm- \nu_\eta d-\psi_\eta]+\frac{1}{N D\bar{K}_\pm}\\
    &K_\pm(d) w_{\eta,j}\bigg),
\end{split}        
\end{equation}
where $\bar{K}_{\pm}$ and $\tilde{K}_{\pm} e^{i \Omega^\eta_{\pm}}$ are the Fourier transforms of the STDP kernels
\begin{align}\label{eq:Kfourier}
&\bar{K}_{\pm}=\int_{-\infty}^{\infty}K_{\pm}(\Delta) d\Delta,\\
&\tilde{K}_{\pm}  e^{i \Omega^\eta_{\pm}}=\int_{-\infty}^{\infty}K_{\pm}(\Delta) e^{-i  \nu_\eta \Delta} d\Delta.
\end{align}
Note that for our specific choice of kernels, $\bar{K}_{\pm}=1$, by construction.

The dynamics of the synaptic weights can be written in terms of the order parameters, $\bar{w}$ and $\tilde{w}$ (see \cref{eq:wbar,eq:wtilda}).
In the continuum limit, \cref{eq:wdot} becomes
\begin{equation}\label{eq:wdotCont}
    \begin{split}
    \frac{\dot{w}_{\eta}(\phi,t)}{\lambda}=&F_{\eta,d}(\phi,t)+\bar{w}_{\eta}(t)F_{\eta,0}(\phi,t)(1+\sigma^2)+\\
    &\tilde{w}_{\eta}(t)F_{\eta,1}(\phi,t)+\bar{w}_{\xi}(t) F_{\eta,0}(\phi,t),
    \end{split}        
\end{equation}
where 
\begin{subequations} \label{eq:Fd01}
	\begin{align}
	\begin{split}
F_{\eta,d}(\phi,t)=&w_{\eta}(\phi,t)\frac{D}{N}\bigg(f_{+}(w_{\eta}(\phi,t))K_+(d)-\\
    & f_{-}(w_{\eta}(\phi,t)) K_-(d)\bigg), 
	\end{split}\\      
F_{\eta,0}(\phi,t)=&D^2 \bigg(\bar{K}_{+}f_{+}(w_{\eta}(\phi,t))   -\bar{K}_{-}f_{-}(w_{\eta}(\phi,t))\bigg),\\
	\begin{split}
	F_{\eta,1}(\phi,t)=&  \frac{\gamma^2}{2}(1+\sigma^2)\bigg(\tilde{K}_+ f_{+}(w_{\eta}(\phi,t)) \cos[\phi-\Omega^\eta_+- \\
	& \nu_\eta d-\psi_\eta]- \tilde{K}_- f_{-}(w_{\eta}(\phi,t)) \cos[\phi-\Omega^\eta_-
	\\ &- \nu_\eta d-
	\psi_\eta]\bigg).
	\end{split}        
\end{align}
\end{subequations}

Integrating \cref{eq:wdotCont} over $\phi$ yields the dynamics
of the order parameters
\begin{equation}\label{eq:wdotbar}
    \begin{split}
    \frac{\dot{\bar{w}}_{\eta}(t)}{\lambda}=&\bar{F}_{\eta,d}(t)+\bar{w}_{\eta}(t)\bar{F}_{\eta,0}(t)(1+\sigma^2)+\\
    &\tilde{w}_{\eta}(t)\bar{F}_{\eta,1}(t)+\bar{w}_{\xi}(t) \bar{F}_{\eta,0}(t),
    \end{split}        
\end{equation}
\begin{equation}\label{eq:wdottilde}
    \begin{split}
    &\frac{1}{\lambda}\frac{d}{dt}(\tilde{w}_{\eta}(t)e^{i\psi_\eta})=\frac{e^{i\psi_\eta}}{\lambda}\big(\dot{\tilde{w}}_\eta(t)+i\tilde{w}_{\eta}(t) \dot{\psi}_\eta (t)\big)=\\
    &\tilde{F}_{\eta,d}(t) e^{i \Phi_{\eta,d}}+\bar{w}_{\eta}(t)(1+\sigma^2)\tilde{F}_{\eta,0}(t) e^{i \Phi_{\eta,0}}+\tilde{w}_\eta\tilde{F}_{\eta,1}(t) \\
    &e^{i \Phi_{\eta,1}}+\bar{w}_{\xi}(t)\tilde{F}_{\eta,0}(t) e^{i \Phi_{\eta,0}},
    \end{split}        
\end{equation}
where 
\begin{subequations} \label{eq:Fd01overphi}
	\begin{align}
	&\bar{F}_{\eta,x}(t)=\int_{-\pi}^{\pi}F_{\eta,x}(\phi,t) \frac{d\phi}{2\pi}, \\
	&\tilde{F}_{\eta,x}(t)e^{i \Phi_{\eta,x}}=\int_{-\pi}^{\pi}e^{i \phi}F_{\eta,x}(\phi,t)  \frac{d\phi}{2\pi}  ,\ x=d,0,1.
	\end{align}
\end{subequations}

Note that in \cref{eq:wdottilde}, $\tilde{w}_2$ does not appear explicitly in the dynamics of $\tilde{w}_1$. This results from the linearity of the post synaptic neuron model we chose, \cref{eq:meanpost}.

\Cref{eq:wdotbar,eq:wdottilde} describe high dimensional coupled non-linear dynamics for the synaptic weights. Studying the development of rhythmic activity is thus not a trivial task. To this end, we took an indirect approach. Below we analyze the homogeneous solution, $w_{\eta,i} = w_\eta$ for all $i$, in which the rhythmic activity does not propagate downstream, and investigate its stability. 
We are interested in investigating the conditions in which the homogeneous solution is unstable, and the STDP dynamics can evolve to a solution that has the capacity to transmit rhythmic information in both channels downstream.

\subparagraph{The homogeneous solution.} 

The symmetry of the STDP dynamics, \cref{eq:wdotCont}, with respect to rotation guarantees the existence of a uniform solution where $w_{\eta,j}(t)=w_\eta^*$ $ \forall j \in \left\lbrace  1,...N \right\rbrace $ and $\tilde {w}_\eta(t)=0$ with $\eta=1,2$.

Solving the fixed point equation for the homogeneous solution yields 

\begin{equation}\label{eq:SteadyStateSolNew}
    \frac{f_-(w^*)}{f_+(w^*)}=\frac{1+X_+}{1+X_-} \equiv \alpha_c,
\end{equation}
where
\begin{equation}\label{eq:Xpm}
    X_\pm \equiv \frac{1}{(2+\sigma^2) N D} K_\pm(d) \geq 0.
\end{equation}
Due to the scaling of $X_\pm$ with $N$, $\alpha_c$ is not expected to be far from $1$. From symmetry  $w_{1}^*=w_{2}^*=w^*$:

\begin{equation}\label{eq:SteadyStateSolHomo}
    w^*=\Bigg(1+ \bigg( \frac{ \alpha}{ \alpha_c}\bigg)^{1/\mu}\Bigg)^{-1}.
\end{equation}

Substituting the homogeneous solution into the post-synaptic firing rate equation, \cref{eq:meanpost} yields

\begin{equation}\label{eq:meanpostHomo}
\langle \rho_{post} \rangle=2 D w^*.
\end{equation}
Thus, in the homogeneous solution, the post-synaptic neuron will fire at a constant rate in time and the rhythmic information will not be relayed downstream.

\subparagraph{Stability analysis of the homogeneous solution.}

Performing standard stability analysis, we consider small fluctuations around the homogeneous fixed point,   $w_{\eta, j}=w^*+\delta {w_{\eta, j}}$, and expand to first order in the fluctuations: 

\begin{equation}\label{eq:StabilityMat}
    \delta \boldsymbol{\dot{w}}=\lambda D^2 \boldsymbol{M} \delta \boldsymbol{w},
\end{equation}
where  $ \boldsymbol{M}$ is the stability matrix.

Using \cref{eq:PotetiationDepression}, the fluctuations can be written as  
\begin{equation}\label{eq:FlucI}
 \delta \dot{w}_{\eta,j}= \delta I_{(\eta,j)}^+-\delta I_{(\eta,j)}^-.
\end{equation}
with
\begin{equation}\label{eq:FlucIpm}
\begin{split}
&\delta I_{(\eta,j)}^\pm=\frac{\partial f_\pm(w^*)}{\partial w_{\eta,j}}D^2(2+\sigma^2)(1+X_\pm)w^* \delta w_{\eta,j}+f_{\pm} D^2\bigg((\\
&1+\sigma^2) \delta \bar{w}_\eta+ \delta \bar{w}_\xi +\frac{\gamma^2}{2}(1+\sigma^2) \tilde{K}_\pm \cos[\phi_{\eta,j}-\Omega_\pm^\eta- \nu_\eta d \\
&-\psi_\eta]\delta \tilde{w}_\eta \bigg)+\frac{f_{\pm}(w^*)}{N}D K_\pm(d)\delta w_{\eta,j}.
\end{split}
\end{equation}
Without loss of generality, taking  $\eta=1$, $\xi=2$  yields
\begin{equation}\label{eq:deltawdot}
\begin{split}
&\delta \dot{w}_{1,j}= -\hat{g}_0 \delta w_{1,j}-\Delta f(w^*)(\delta \bar{w}_1+\delta \bar{w}_2)-\sigma^2 \Delta f(w^*)\delta \bar{w}_1\\
&+\gamma^2(1+\sigma^2)\big(f_+(w^*)\tilde{K}_+(\nu_1) \cos[\phi_{1,j}-\Omega_+^1- \nu_1 d-\psi_1 ] \\
& -f_-(w^*)\tilde{K}_-(\nu_1) \cos[\phi_{1,j}-\Omega_-^1- \nu_1 d-\psi_1]\big)\delta \tilde{w}_1.
\end{split}
\end{equation}
In the homogeneous fixed point, \cref{eq:SteadyStateSolNew}:
%
\begin{equation}\label{eq:g0New}
\begin{split}
\hat{g_0}&=(2+\sigma^2) \bigg( \alpha \mu(1+X_-)\frac{w^{*\mu}}{1-w^*}+f_+(w^*)-f_-(w^*) \bigg)\\
&=g_0-(2+\sigma^2) \Delta f(w^*),
\end{split}
\end{equation}
where 

\begin{subequations} \label{eq:g0andDeltaf}
	\begin{align}
	&g_0\equiv \alpha \mu (2+\sigma^2) (1+X_-)\frac{w^{*\mu}}{1-w^*}\\
	&\Delta f(w)\equiv f_-(w)-f_+(w).
	\end{align}
\end{subequations}

Studying \cref{eq:deltawdot}, the stability analysis around the homogeneous fixed point yields four prominent eigenvectors. Two are in the subspace of the uniform directions of the two populations, and two are in  directions of the first Fourier modes of each population.
As the uniform modes of fluctuations, $\delta \bar{\boldsymbol{w}}^\top=( \delta \bar{w_{1}}  ,\  \delta \bar{w_{2}})$, span an invariant subspace of the stability matrix,  $ \boldsymbol{M}$, we can study the restricted matrix, $\bar{\boldsymbol{M}}$, defined by:
\begin{equation}\label{eq:stabilitydecompostionSubMatrix}
\delta \boldsymbol{\dot{\bar{w}}}=\lambda D^2 \bar{\boldsymbol{M}} \delta  \boldsymbol{\bar{w}}. 
\end{equation}

\begin{figure*}[tb]
	\centering
	\begin{subfigure}[t]{0.008\textwidth}
		\textbf{(a)} 
	\end{subfigure}
	\begin{subfigure}[t]{0.45\textwidth}  \vspace{0.0005\textwidth}
		\includegraphics[width=\linewidth, valign=t]{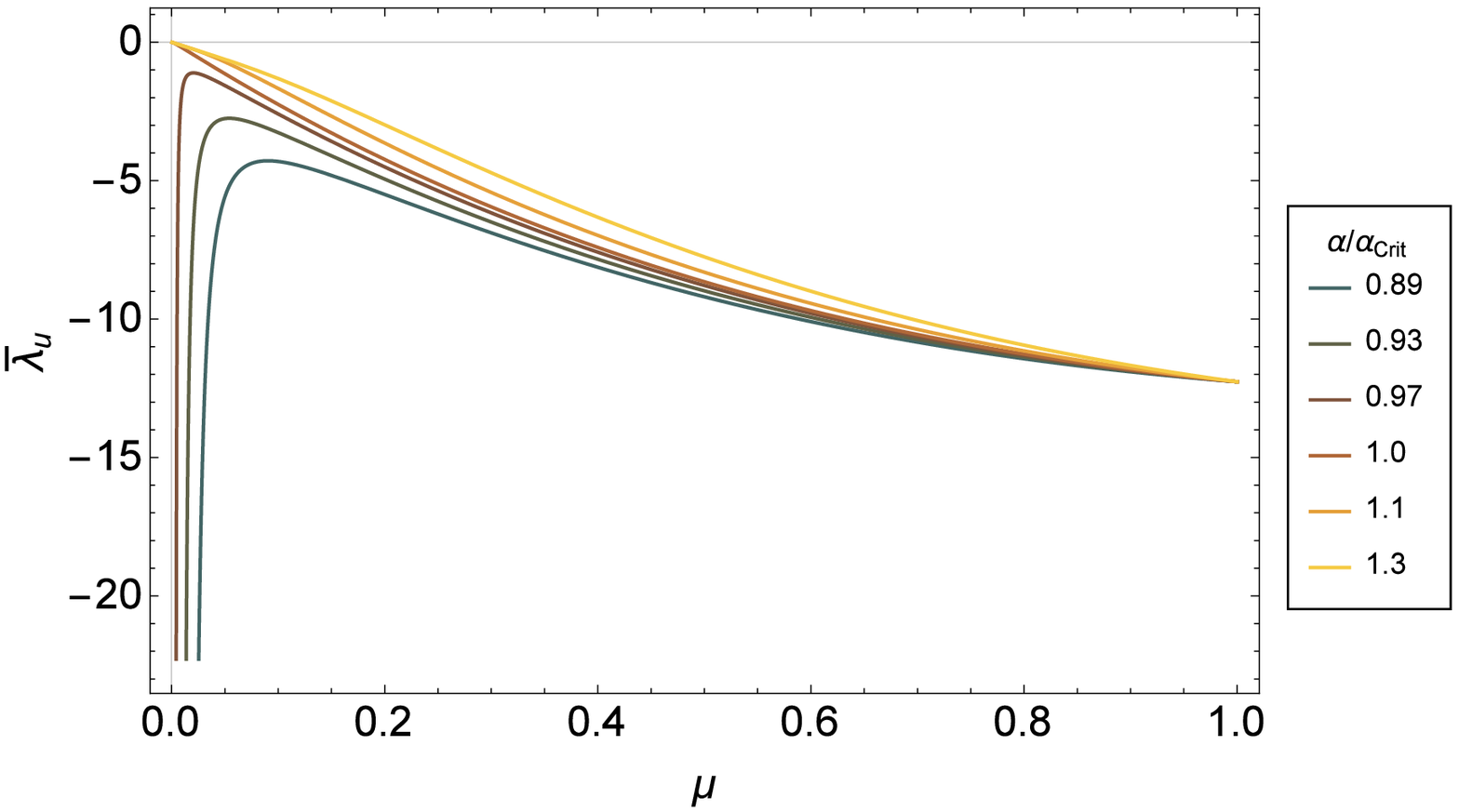}\\[3pt]
		\caption{}
		\label{fig:LambdaUniformAsymmetric}
	\end{subfigure}\hfill \vspace{0.02\textwidth}
	\begin{subfigure}[t]{0.008\textwidth}
		\textbf{(b)}
	\end{subfigure}
	\begin{subfigure}[t]{0.45\textwidth}  \vspace{0.0005\textwidth} 	
		\includegraphics[width=\linewidth, valign=t]{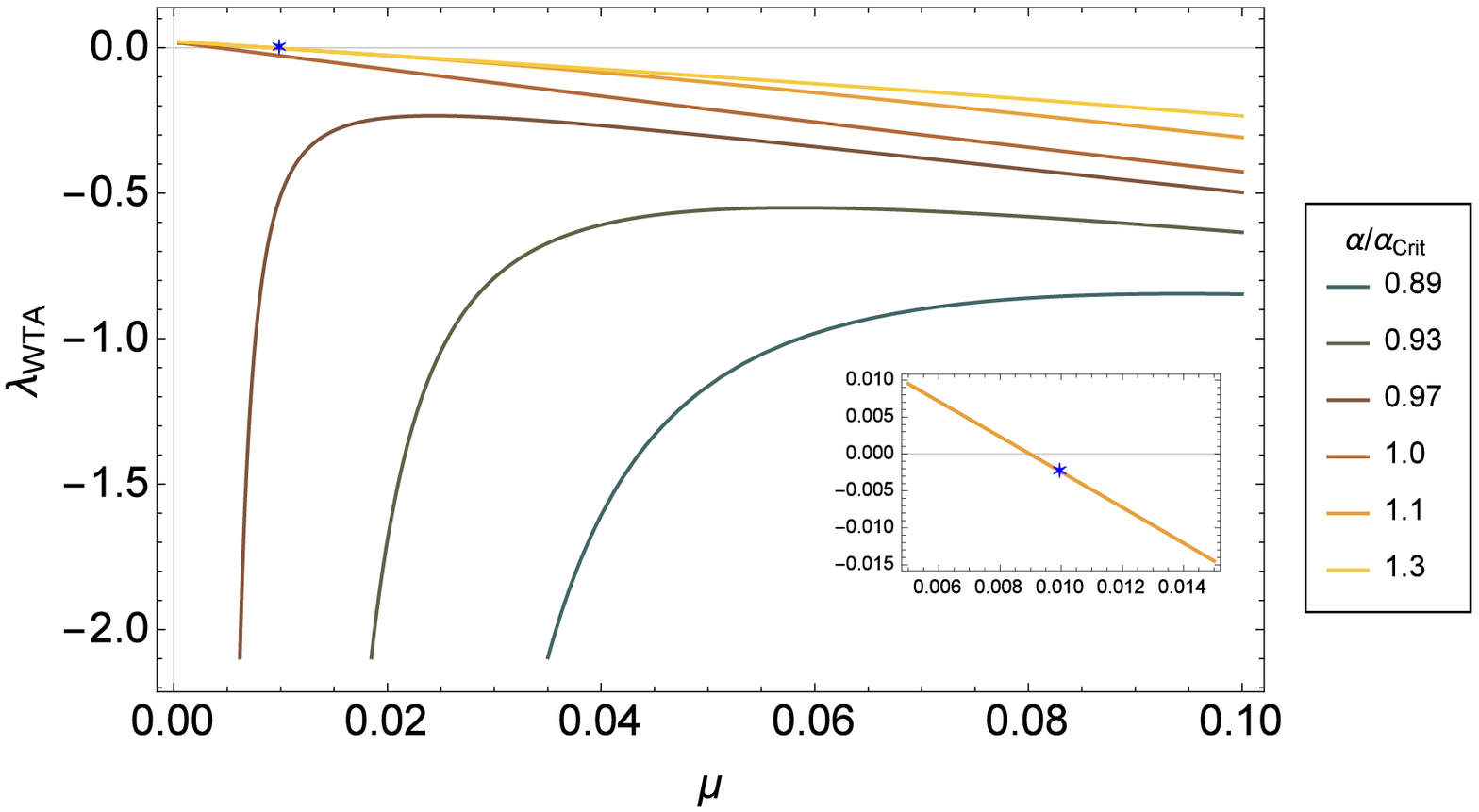}\\[3pt]
		\caption{}
		\label{fig:LambdaWTAAsymmetric}
	\end{subfigure}\hfill \vspace{0.02\textwidth}

	\begin{subfigure}[t]{0.008\textwidth}
		\textbf{(c)}
	\end{subfigure}
	\begin{subfigure}[t]{0.45\textwidth}  \vspace{0.0005\textwidth} 	
		\includegraphics[width=\linewidth, valign=t]{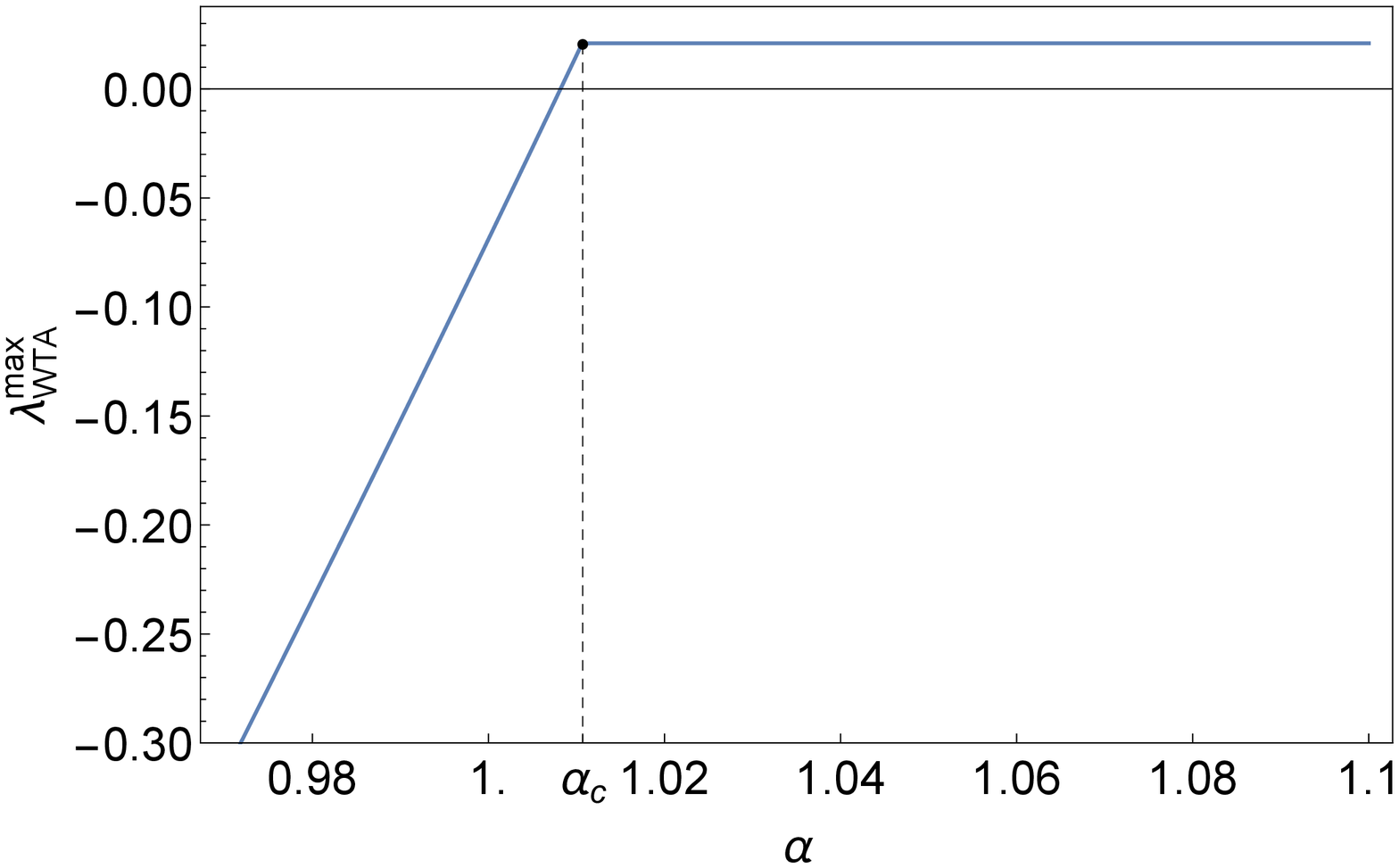}\\[3pt]
		\caption{}
		\label{fig:lambdaMax}
	\end{subfigure}\hfill \vspace{0.02\textwidth}
	\caption{The eigenvalues of the uniform and winner-take-all modes for the asymmetric STDP rule, \cref{eq:kernel}. (a) The uniform eigenvalue $\bar{\lambda}_\text{u}$ is shown as a function of $\mu$, for different values of $\alpha/\alpha_\text{c}$  which are depicted by color. (b)  The eigenvalue of the competitive winner-take-all mode, $\lambda_\text{WTA}$, is shown as a function of $\mu$, for different values of $\alpha/\alpha_\text{c}$ as depicted by color. (Inset) Enlarged section of the figure, showing the eigenvalue corresponding to the choice of parameters in our simulations below, \cref{subsec:Numerical} -depicted by a blue star. (c) The maximal value of $\lambda_\text{WTA}$ in the interval $\mu \in (0,1)$ is shown as a function of $\alpha$. Note that for $\alpha \geq \alpha_\text{c}$, $\lambda^\text{max}_\text{WTA}$ is obtained on the boundary $\mu = 0$. Unless stated otherwise in the legends, the parameters used in these figures are $\sigma=0.6$, $D=10  \text{spikes}/\text{sec}$, $N=120$, $\tau_-=50 \text{msec}$, $\tau_+=20 \text{msec}$ and $d=10\text{msec}$.}  
	\label{fig:lambdaTauUniformWTA}
\end{figure*}

The matrix $ \bar{\boldsymbol{M}}$ has two eigenvectors, $\boldsymbol{v}_\text{u}^\top=(1,1)$ and $\boldsymbol{v}_{\text{WTA}}^\top=(1,-1)$. The corresponding eigenvalues are 
 \begin{subequations}
 	\begin{align}\label{eq:eigenvaluesuniformdirection}
 	& \bar{\lambda}_\text{u}= - g_0
 	\\
 \begin{split} \label{eq:LambdaWTA}
 	&\lambda_{\text{WTA}}=
 	 \bar{\lambda}_\text{u}+2 \Delta  f(w^*).
 	  \end{split}
 	\end{align}
 \end{subequations}
The first eigenvector represents the uniform mode of fluctuations and its eigenvalue is always negative, \cref{fig:LambdaUniformAsymmetric}; hence, the homogeneous solution is always stable with respect to uniform fluctuations. Furthermore, $\bar{\lambda}_\text{u}$ serves as a stabilizing term in other modes of fluctuations. We distinguish two regimes: $\alpha < \alpha_c$ and $\alpha > \alpha_c$. For $\alpha < \alpha_c$, $ \lim_{ \mu \rightarrow 0+} \bar{\lambda}_\text{u} = - \infty  $, and the uniform solution is expected to remain stable. For $\alpha > \alpha_c$, $ \lim_{ \mu \rightarrow 0+} \bar{\lambda}_\text{u} = 0 $, and structure may emerge for sufficiently low values of $\mu$; see legend color map in \cref{fig:LambdaUniformAsymmetric}.  

The second eigenvalue represents a winner-take-all like mode of fluctuations, in which synapses in one population suppress the other. Clearly, a winner-take-all competition will prevent multiplexing. For $\alpha > \alpha_c$, $ \lim_{ \mu \rightarrow 0+} \lambda_\text{WTA} = 2 (\alpha_c - 1) $. For the temporally asymmetric learning rule, \cref{eq:kernel}, $X_- = 0$; consequently $\alpha_c >1$. In the temporally symmetric difference of Gaussians STDP model, \cref{eq:kernelSymmetric}, $\alpha_c >1$ if and only if $\tau_+ < \tau_-$, which is the typical case. 
For $\alpha < \alpha_c$ in the limit of small $\mu$ the divergence of $\bar{\lambda}_\text{u}$ stabilizes fluctuations in this mode. In this case ($\alpha < \alpha_c$), $\lambda_\text{WTA}$ reaches its maximum at an intermediate value of $\mu \in (0,1)$, \Cref{fig:LambdaWTAAsymmetric}. For a small range of $\alpha < \alpha_c$, $\alpha \approx \alpha_c$ this maximum can be positive, see \cref{fig:lambdaMax}.   
\Cref{fig:LambdaWTAAsymmetric} depicts $\lambda_\text{WTA}$ as a function of $\mu$ for different values of $\alpha$ as shown by color. Note that $\lambda_\text{WTA}$ depends on the temporal structure of the STDP rule solely via the value of $\alpha_c$ and $X_-$; however, its sign is independent of $X_-$. As can seen in the figure, for $\alpha > \alpha_c > 1$, $\lambda_\text{WTA}$ is a decreasing function of $\mu$ and the homogeneous solution loses stability in the competitive winner-take-all direction in the limit of small $\mu$.  Note that these eigenvalues are not completely identical  in the asymmetric and symmetric cases due to the fact that $X_-^\text{Symmetric} \ne X_-^\text{Asymmetric}=0$. However,  $X_-^\text{Symmetric} \sim 10^{-3}$; hence, both rules are qualitatively similar. 




The rhythmic modes are eigenvectors of the stability matrix  $\boldsymbol{M}$ with eigenvalues
 \begin{align}\label{eq:eigenvaluesOscillatory}
 \tilde{\lambda}_{\nu_\eta}= & \bar{\lambda}_\text{u} +  (2+\sigma^2) \Delta f(w^*)  +\\ 
 \nonumber
 & \gamma^2 (1+\sigma^2) f_+(w^*) \tilde{Q}
 \\
 \tilde{Q} =&   \tilde{K}_+(\nu_\eta) \cos[\Omega_+^\eta+    \nu_\eta d]- \\ 
 \nonumber
  & \alpha_c \tilde{K}_-(\nu_\eta) \cos[\Omega_-^\eta+ \nu_\eta d] ,
 \end{align}
where $\eta=1, 2$.  The first two terms which do not depend on the temporal structure of learning rule or on the frequency, contain the stabilizing term $\bar{\lambda}_\text{u} \leq 0$, and their dependence on $\alpha$ and $\mu$ is similar to that of $\lambda_{\text{WTA}}$; compare with \cref{eq:LambdaWTA}. The last term depends on the real part of the Fourier transform of the delayed STDP rule at the specific frequency of oscillations, $\nu_\eta$. This last term can  destabilize the system in a direction that will enable the propagation of rhythmic activity downstream while keeping the competitive WTA mode stable depending on the interplay between the rhythmic activity and the temporal structure of the STDP rule. For $\tilde{Q}>0$, $\tilde{\lambda}_{\nu_\eta}$ is an increasing function of  the modulation to the mean ratio $\gamma$. If in addition $\alpha > \alpha_c >1$ then $\tilde{\lambda}_{\nu_\eta}$ is an increasing function of $\sigma$ as well. 

In the low frequency limit,  $ \lim_{ \nu \rightarrow 0} \tilde{Q}  = 1- \alpha_c$, and  depends on the characteristic timescales of $\tau_+$, $\tau_-$, and $d$ only via $\alpha_c$. 
For large frequencies $ \lim_{ \nu \rightarrow \infty} \tilde{Q}  = 0$.  
In this limit the STDP dynamics loses its sensitivity to the rhythmic activity. Consequently, the resultant modulation of the synaptic weights profile, $\tilde{w}$, will become negligible; hence, effectively rhythmic information will not propagate downstream even if the rhythmic eigenvalue is unstable \cite{luz2016oscillations}. Thus, the intermediate frequency region is the most relevant for multiplexing.

\begin{figure*}[tb]
	\centering
	\begin{subfigure}[t]{0.008\textwidth}
		\textbf{(a)} 
	\end{subfigure}
	\begin{subfigure}[t]{0.45\textwidth}  \vspace{0.0005\textwidth}
		\includegraphics[width=\linewidth, valign=t]{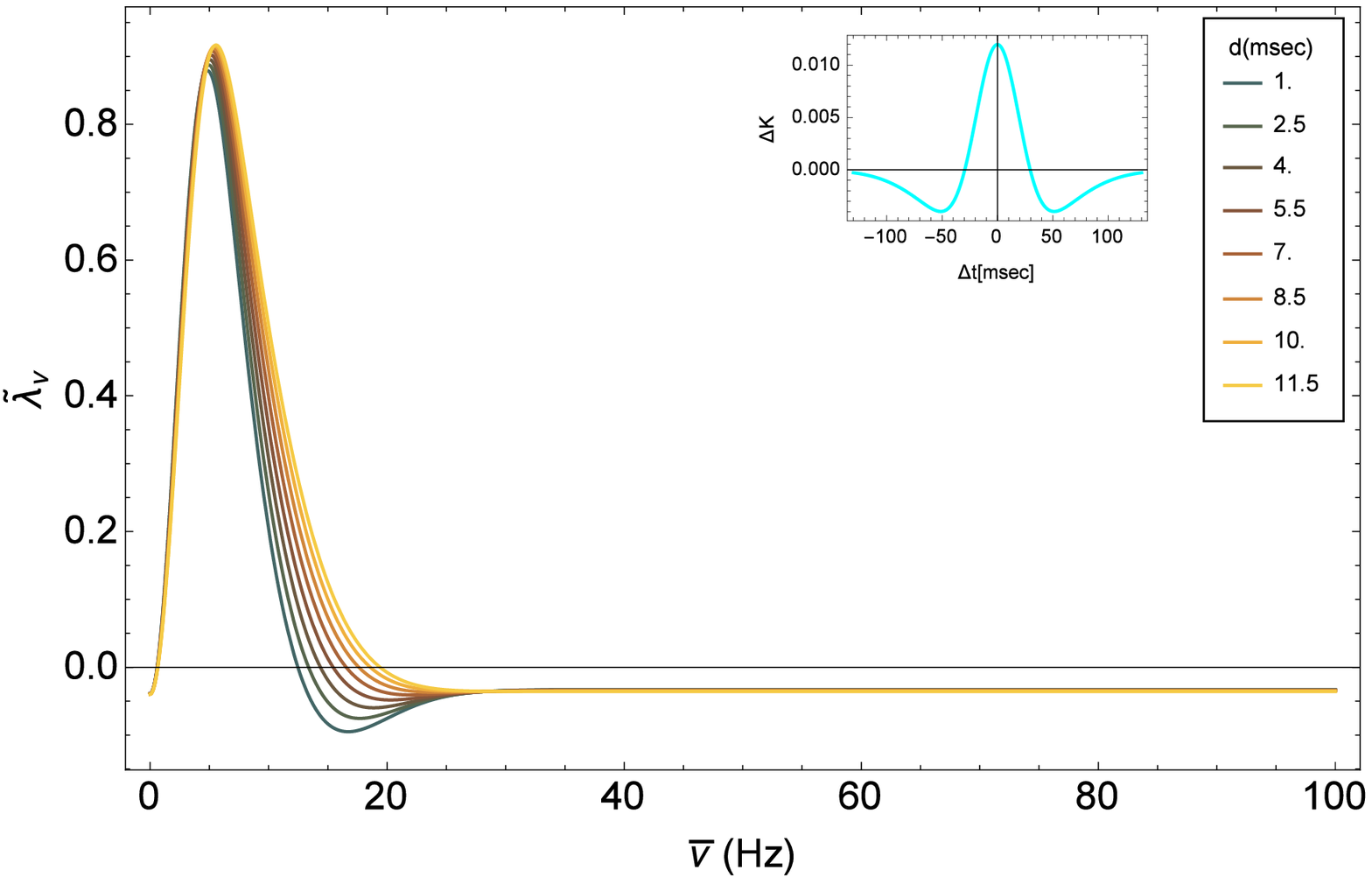}\\[3pt]
		\caption{}
		\label{fig:lambdatildeSymmDelay}
	\end{subfigure}\hfill 
	\begin{subfigure}[t]{0.008\textwidth}
		\textbf{(b)}
	\end{subfigure}
	\begin{subfigure}[t]{0.45\textwidth}  \vspace{0.0005\textwidth} 	
		\includegraphics[width=\linewidth, valign=t]{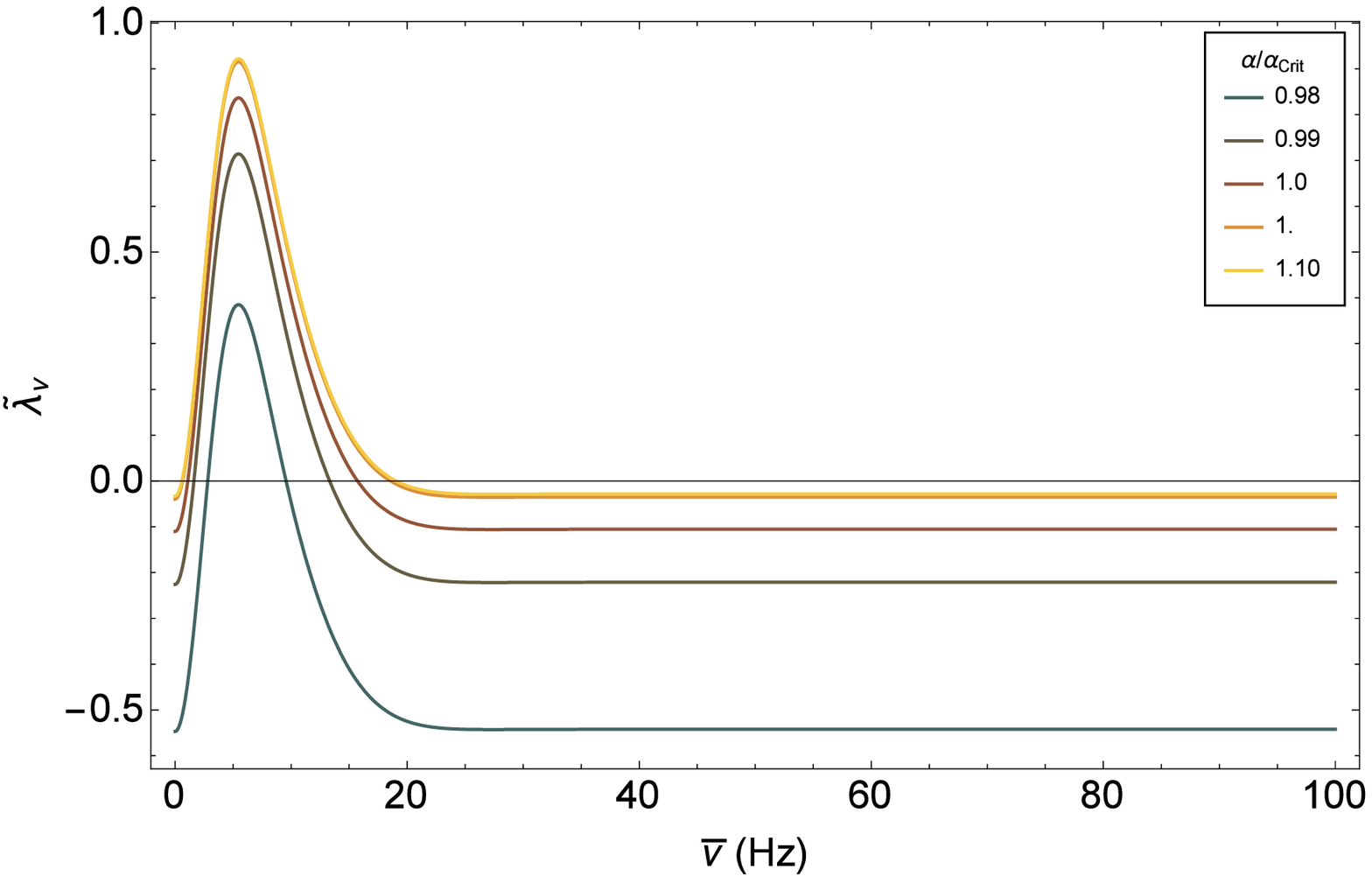}\\[3pt]
		\caption{}
		\label{fig:lambdatildeSymmAlpha}
	\end{subfigure}\hfill \vspace{0.02\textwidth}

		\begin{subfigure}[t]{0.008\textwidth}
		\textbf{(c)} 
	\end{subfigure}
	\begin{subfigure}[t]{0.45\textwidth}  \vspace{0.0005\textwidth}
		\includegraphics[width=\linewidth, valign=t]{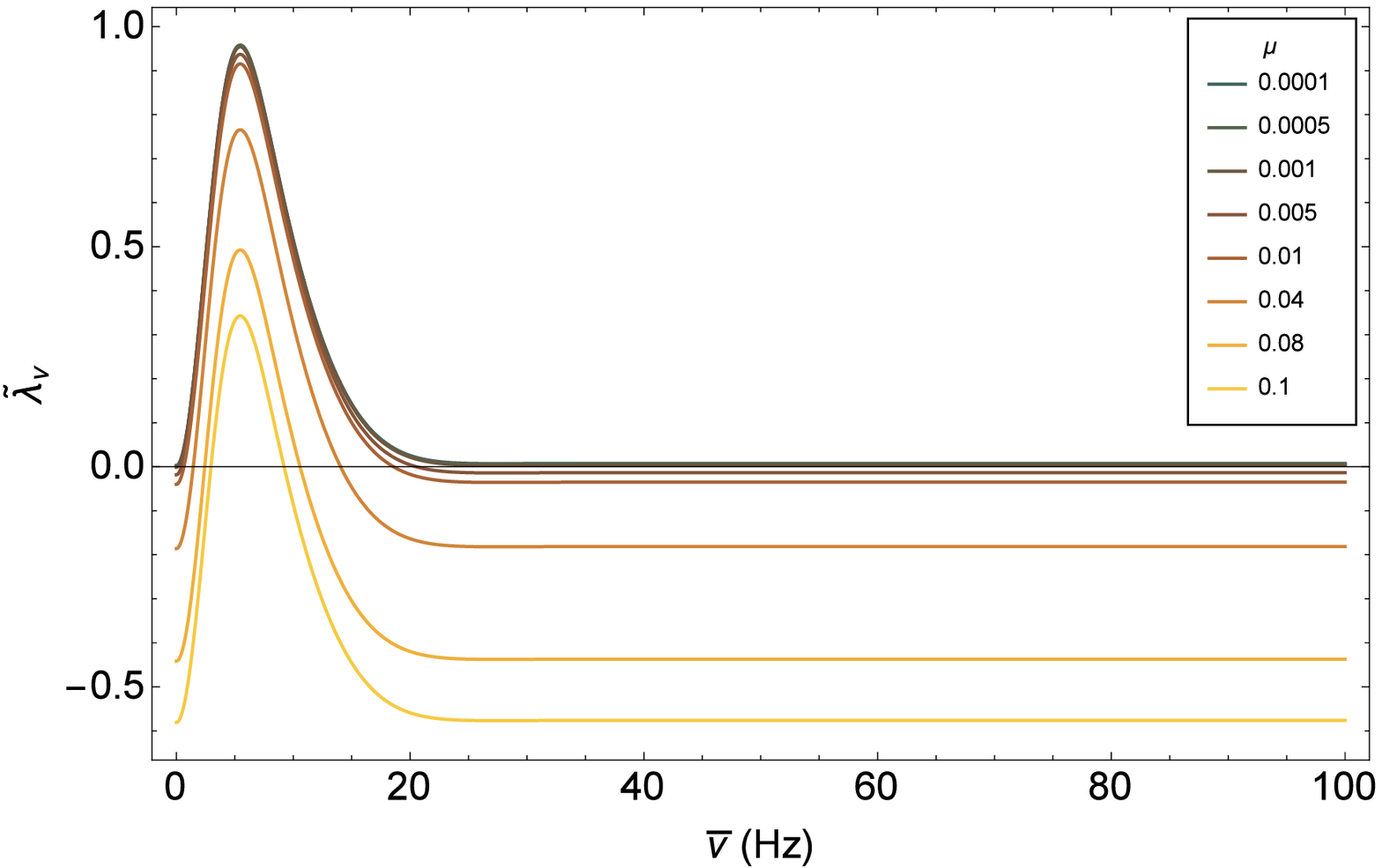}\\[3pt]
		\caption{}
		\label{fig:lambdatildeSymmMiu}
	\end{subfigure}\hfill 
	\begin{subfigure}[t]{0.008\textwidth}
		\textbf{(d)}
	\end{subfigure}
	\begin{subfigure}[t]{0.45\textwidth}  \vspace{0.0005\textwidth} 	
		\includegraphics[width=\linewidth, valign=t]{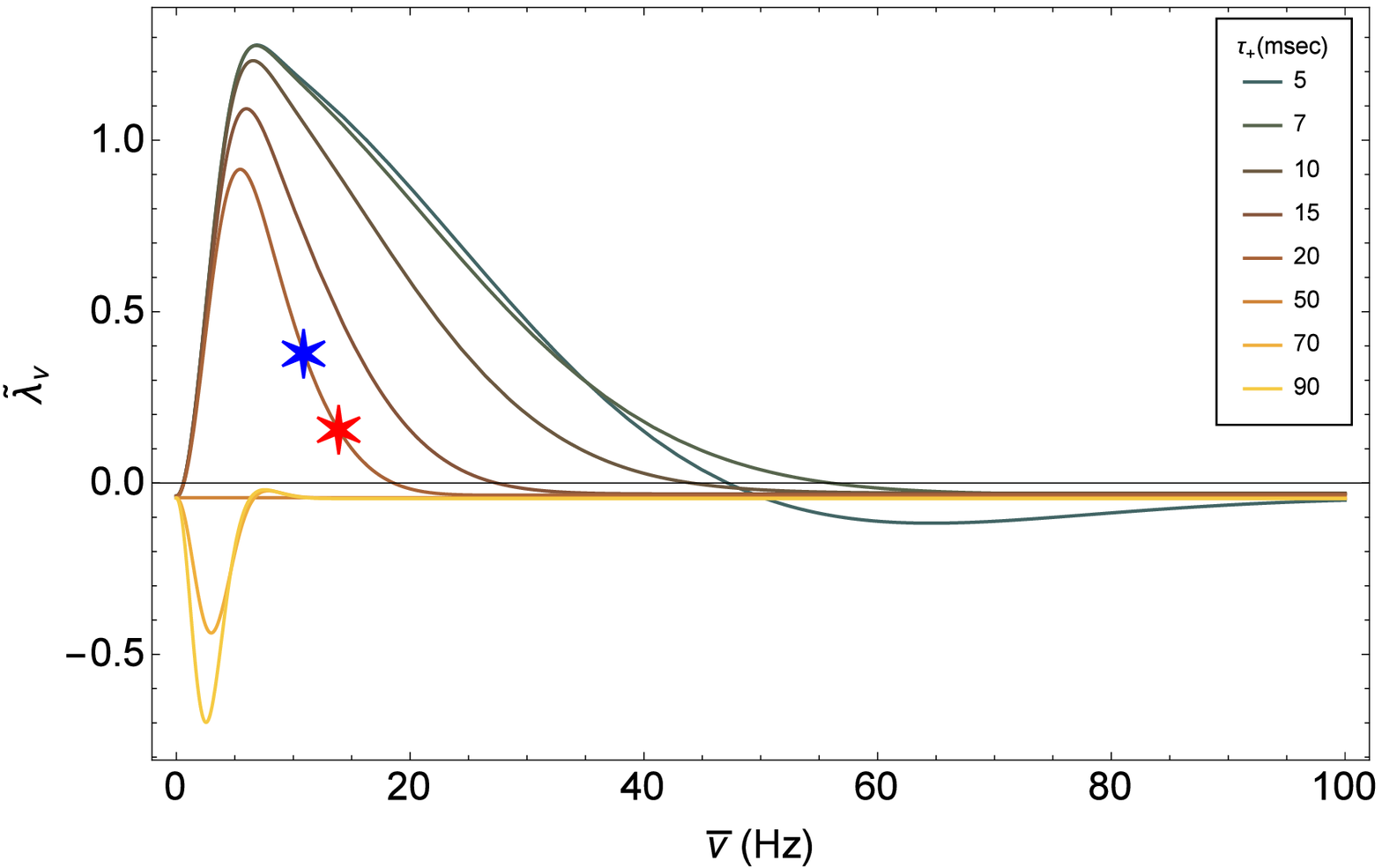}\\[3pt]
		\caption{}
		\label{fig:lambdatildeSymmTau}
	\end{subfigure}\hfill \vspace{0.02\textwidth}
	\caption{The rhythmic eigenvalue, $\tilde{\lambda}_\nu$, in the case of the symmetric learning rule, \cref{eq:kernelSymmetric}. (a) $\tilde{\lambda}_\nu$ is shown  as a function of frequency, $\bar{\nu} = \nu/(2 \pi)$, for different values of $d$ as depicted by color. (b) $\tilde{\lambda}_\nu$ is shown  as a function of frequency for different values of $\alpha/\alpha_\text{c}$ as depicted by color. (c) $\tilde{\lambda}_\nu$  as a function of frequency for different values of $\mu$ - by color. (d) $\tilde{\lambda}_\nu$ as a function of frequency for different values of $\tau_+$ - by color, with $\tau_-=50 \text{msec}$. The blue (11Hz) and red (14Hz) stars show the eigenvalue corresponding to the parameters of the simulations, in  \cref{subsec:Numerical}. Unless stated otherwise in the legends, the parameters used to produce these figures are $\gamma=1$, $\sigma=0.6$, $D=10 \text{spikes}/\text{sec}$, $N=120$, $\tau_-=50 \text{msec}$, $\tau_+=20\text{msec}$, $\mu=0.01$, $\alpha=1.1$ and $d=10\text{msec}$.}
	\label{fig:lambdaTauSymm}
\end{figure*}

In the case of the temporally symmetric kernel, \cref{eq:kernelSymmetric}, the value of $\tilde{Q}$ is given by
\begin{align}\label{eq:eigenvaluesOscillatorySymmetric}
&\tilde{Q} = \cos[ 	\nu_\eta d] \Big( e^{-\frac{1}{2} (\nu_\eta \tau_+)^2}- \alpha_c e^{-\frac{1}{2} (\nu_\eta \tau_-)^2} \Big),
\end{align}
where $\Omega_\pm^\eta = 0$. \Cref{fig:lambdatildeSymmDelay} shows the rhythmic eigenvalue, $\tilde{\lambda}_{\nu}$, as a function of the oscillation frequency, $\bar{\nu} \equiv \nu/(2\pi)$, for different values of the delay, $d$ as depicted by color. Since typically, $\tau_+ < \tau_- $, for finite $\nu$, $\tilde{K}_+(\nu) > \tilde{K}_-(\nu)$. Consequently, $\tilde{Q}$ will be dominated by the potentiation term, $\cos[ 	\nu_\eta d]  e^{-\frac{1}{2} (\nu_\eta \tau_+)^2}$, except for the very low frequency range of $\nu  \lesssim 1/ \tau_-$. Typical values for the delay, $d$, are 1-10$ms$, whereas typical values for the characteristic timescales for the STDP, $\tau_{\pm}$, are tens of $ms$. As a result, the specific value of the delay, $d$, does not affect the rhythmic eigenvalue much, and the system becomes unstable in the rhythmic direction for $ 1/\tau_- \lesssim \nu \lesssim 1/\tau_+$. 

Increasing the relative strength of the depression, $\alpha$, strengthens the stabilizing term $\bar{\lambda}_\text{u}$; however, $\Delta f(w^*)$ scales approximately linearly with $\alpha$ such that the rhythmic eigenvalue is elevated, causing the frequency range in which $\tilde{\lambda}_{\nu}>0$ to widen, \cref{fig:lambdatildeSymmAlpha}. Similarly, for $\alpha > \alpha_c >1$, increasing $\mu$ strengthens $\bar{\lambda}_\text{u}$ and reduces the frequency range in which $\tilde{\lambda}_{\nu}>0$, \cref{fig:lambdatildeSymmMiu}. Decreasing the characteristic timescale of potentiation, $\tau_+$ increases the frequency region with an unstable rhythmic eigenvalue; however, when $\tau_+$ becomes comparable to the delay, $d$, the oscillatory nature of $\tilde{\lambda}_{\nu}$ in $\nu$ is revealed, \cref{fig:lambdatildeSymmTau}.

In the case of the temporally asymmetric kernel, \cref{eq:kernel}, the value of $ \tilde{Q}$ is given by
\begin{equation}\label{eq:eigenvaluesOscillatoryAsymmetric}
\tilde{Q}^ = \frac{ \cos[\Omega_+^\eta+
 \nu_\eta d]}{\sqrt{1+ (\nu_\eta \tau_+)^2 }} - 
  \alpha_c \frac{ \cos[\Omega_-^\eta+
	\nu_\eta d]}{\sqrt{1+ (\nu_\eta \tau_-)^2 }}  ,
\end{equation}
with $\Omega_\pm^\eta=\mp\arctan( \nu_\eta \tau_{\pm})$. 

\begin{figure*}[tb]
	\centering
	\begin{subfigure}[t]{0.008\textwidth}
		\textbf{(a)} 
	\end{subfigure}
	\begin{subfigure}[t]{0.45\textwidth}  \vspace{0.0005\textwidth}
		\includegraphics[width=\linewidth, valign=t]{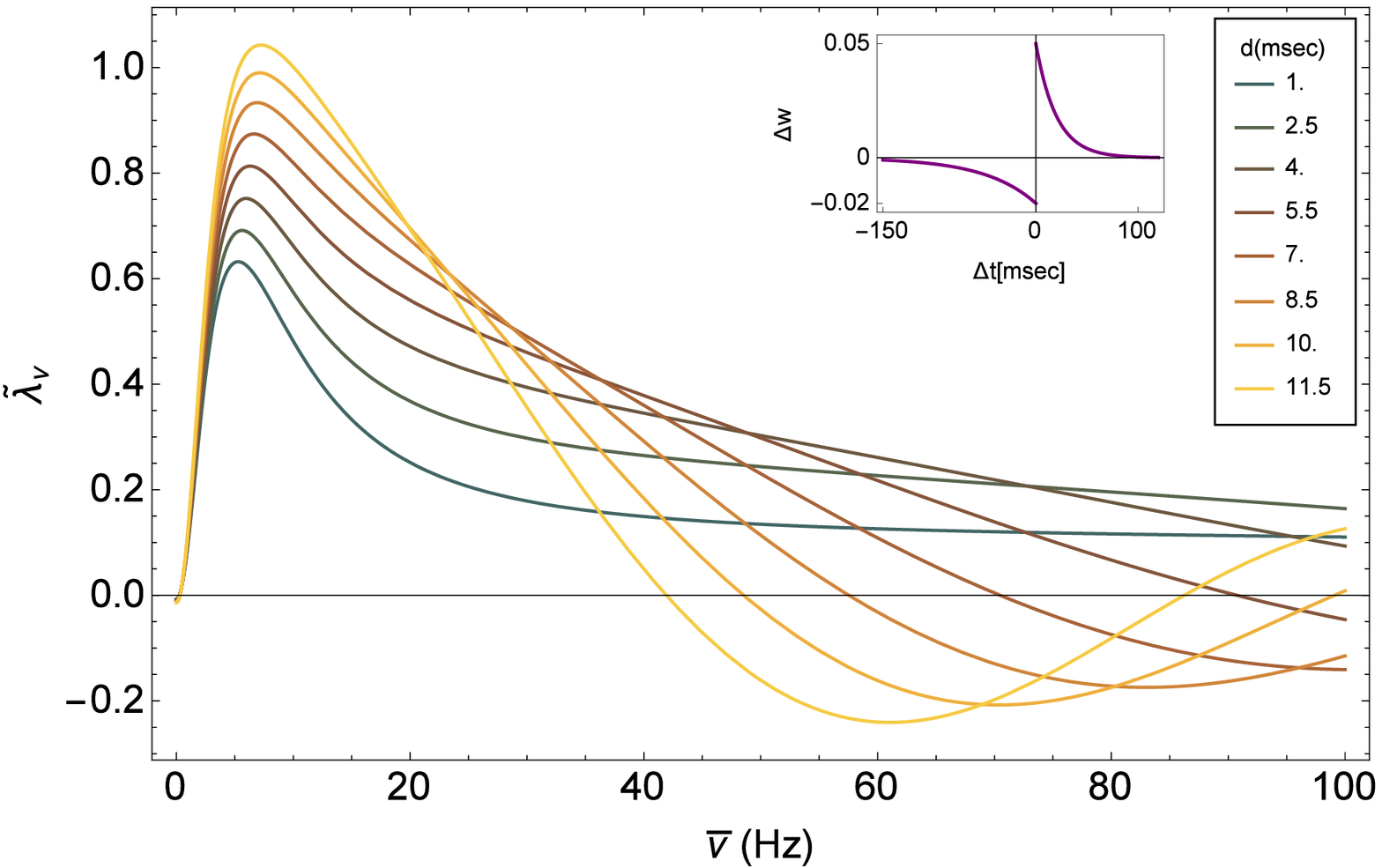}\\[3pt]
		\caption{}
		\label{fig:lambdatildeAsymmDelayFrame}
	\end{subfigure}\hfill 
	\begin{subfigure}[t]{0.008\textwidth}
		\textbf{(b)}
	\end{subfigure}
	\begin{subfigure}[t]{0.45\textwidth}  \vspace{0.0005\textwidth} 	
		\includegraphics[width=\linewidth, valign=t]{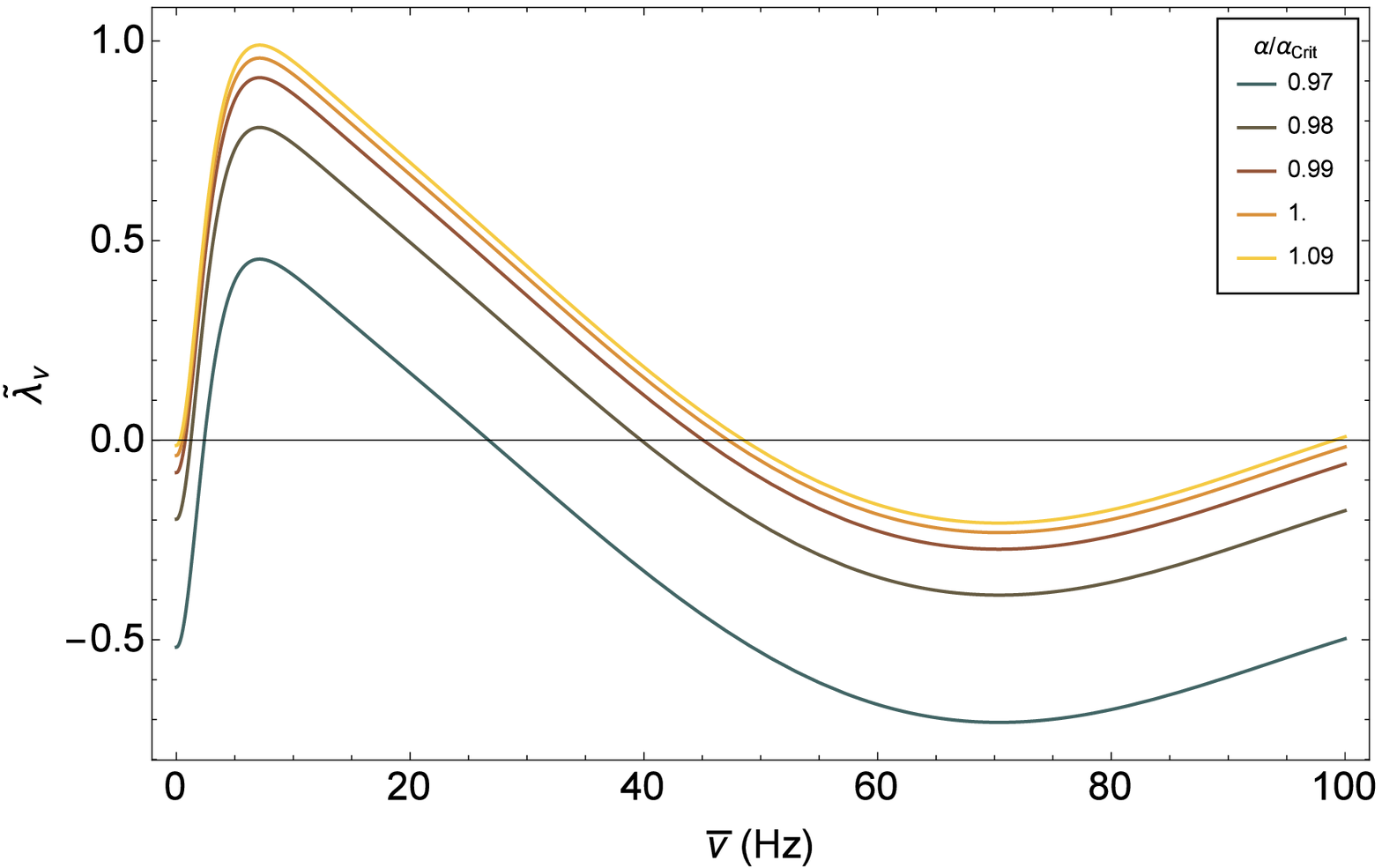}\\[3pt]
		\caption{}
		\label{fig:lambdatildeAsymmAlpha}
	\end{subfigure}\hfill \vspace{0.02\textwidth}

		\begin{subfigure}[t]{0.008\textwidth}
		\textbf{(c)} 
	\end{subfigure}
	\begin{subfigure}[t]{0.45\textwidth}  \vspace{0.0005\textwidth}
		\includegraphics[width=\linewidth, valign=t]{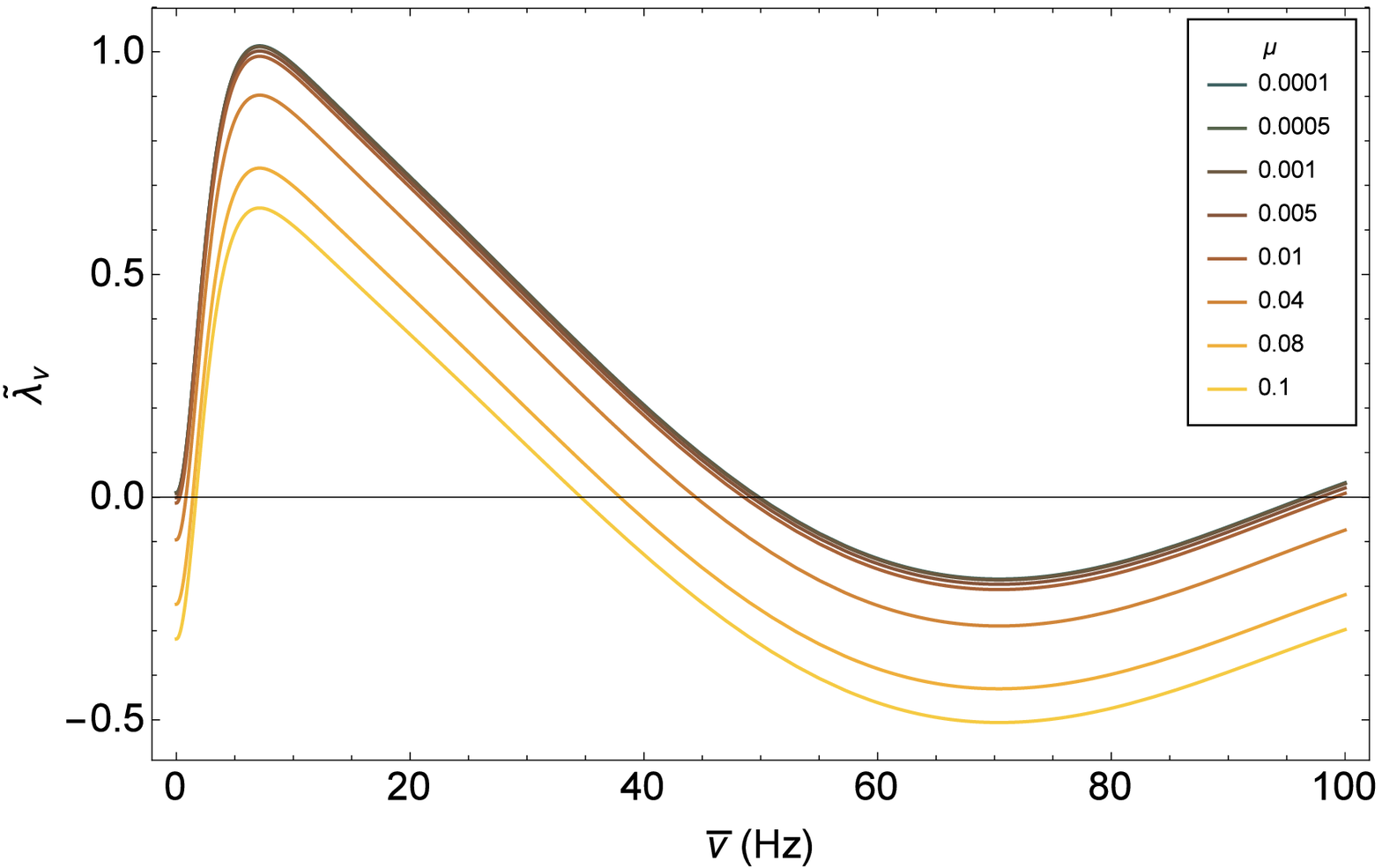}\\[3pt]
		\caption{}
		\label{fig:lambdatildeAsymmMiu}
	\end{subfigure}\hfill 
	\begin{subfigure}[t]{0.008\textwidth}
		\textbf{(d)}
	\end{subfigure}
	\begin{subfigure}[t]{0.45\textwidth}  \vspace{0.0005\textwidth} 	
		\includegraphics[width=\linewidth, valign=t]{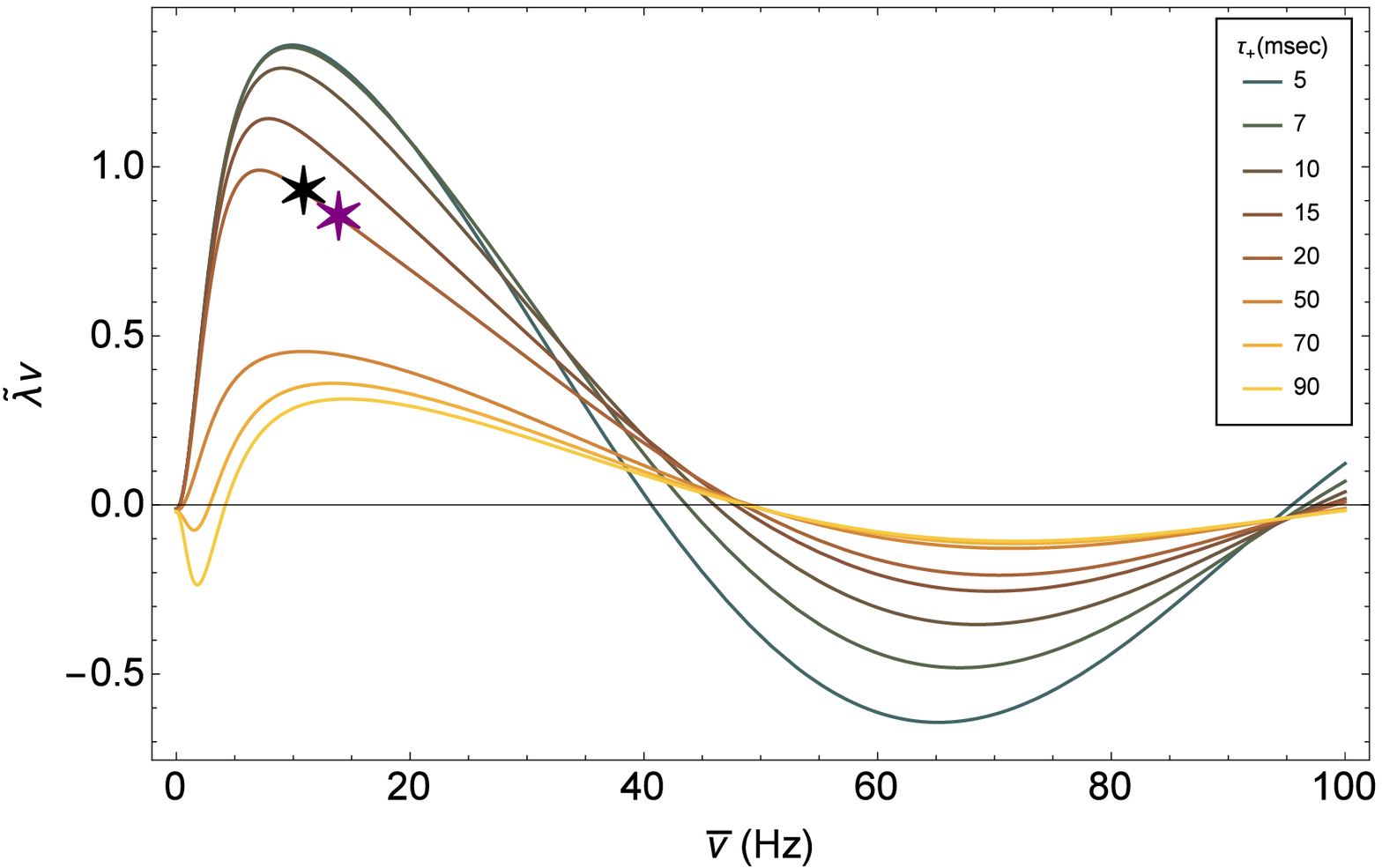}\\[3pt]
		\caption{}
		\label{fig:lambdatildeAsymmTau}
	\end{subfigure}\hfill \vspace{0.02\textwidth}
	\caption{The rhythmic eigenvalue, $\tilde{\lambda}_\nu$, in the case of the asymmetric learning rule, \cref{eq:kernel}. (a) $\tilde{\lambda}_\nu$ is shown as a function of the frequency, $\bar{\nu} = \nu/(2 \pi)$, for different values of $d$ as depicted by color. (b) $\tilde{\lambda}_\nu$  is shown as a function of the frequency, for different values of $\alpha/\alpha_\text{c}$ as depicted by color. (c) $\tilde{\lambda}_\nu$ as a function of the frequency, for different values of $\mu$ - by color. (d) $\tilde{\lambda}_\nu$ as a function of frequency, for different values of $\tau_+$ - by color,  with $\tau_-=50 \text{msec}$. The black (11Hz) and purple (14Hz) stars show the eigenvalue with the corresponding parameters to the simulations in \cref{subsec:Numerical}.  Unless stated otherwise in the legends, the parameters used to produce these figures are $\gamma=1$, $\sigma=0.6$, $D=10 \text{spikes}/\text{sec}$, $N=120$, $\tau_-=50 \text{msec}$, $\tau_+=20\text{msec}$, $\mu=0.01$, $\alpha=1.1$ and $d=10\text{msec}$.}
	\label{fig:lambdaTauAsymm}
\end{figure*}

The main difference between the temporally symmetric and the asymmetric rules is that due to the discontinuity of the asymmetric STDP kernel, $\tilde{K}_{\pm}$ decay algebraically rather than exponentially fast with  $\nu$. As a result, the phase $\cos[\Omega_{\pm}^\eta+ \nu_\eta d]$, plays a more central role in controlling the stability of the rhythmic eigenvalue. 
As above, since typically, $\tau_+ < \tau_- $, then  $\tilde{K}_+(\nu) > \tilde{K}_-(\nu)$. \Cref{fig:lambdatildeAsymmDelayFrame} shows the rhythmic eigenvalue, $\tilde{\lambda}_{\nu}$, for different values of $d$. The dashed vertical lines depict the frequencies at which the potentiation term changes its sign, $\Omega_{+}^\eta+ \nu^* d = \pi/2$. As can be seen from the figure, for this choice of parameters the upper cutoff of the central frequency range in which the rhythmic eigenvalue is unstable is dominated by $\nu^*$, which is governed by the delay.  

The  effects of parameters $\alpha$ and $\mu$ show similar trends as for the symmetric STDP rule. Specifically, increasing $\mu$ or decreasing $\alpha$, in general, shrinks the region in which fluctuations in the rhythmic direction are unstable, \cref{fig:lambdatildeAsymmAlpha,fig:lambdatildeAsymmMiu}. Increasing the characteristic timescale of potentiation, $\tau_+$ beyond that of the depression, makes the depression term, $\tilde{K}_-(\nu_\eta) \cos[\Omega_-^\eta+ \nu_\eta d]$, more dominant, \cref{fig:lambdatildeAsymmTau}. In this case the lower frequency cutoff will be dominated by the change of sign in the depression term; i.e., by the angular frequency $\nu^*$, such that $\Omega_{-}^\eta+ \nu^* d = \pi/2$, shown by the light brown curve ($\tau_+= 50 \text{msec}$) in \cref{fig:lambdatildeAsymmTau}.

From the above analysis, one would expect that multiplexing will emerge via STDP if both rhythmic eigenvalues are unstable and the winner-take-all eigenvalue is stable. However, this intuition relies on two major factors. The first is that the, stability analysis is local and can only describe the dynamics near the homogeneous fixed point. The second is the choice of a linear neuron model, \cref{eq:meanpost}, which resulted in the lack of explicit interaction term between the two rhythms $\tilde{w}_1$ and  $\tilde{w}_2$ in \cref{eq:wdottilde}. Both of these issues are addressed below by studying the STDP dynamics using a conductance based Hodgkin-Huxley type numerical model of a spiking post-synaptic neuron.

\subsection{Conductance based neuron model}  \label{subsec:Numerical}

In our numerical analysis we used the conductance based model described in Shriki et al. \cite{shriki2003rate} for the post-synaptic neuron. This choice was motivated by the ability to control the linearity of the neuron's response to its inputs. Often the response of a neuron is quantified using an f-I curve, which maps the frequency (f) of the neuronal spiking response to a certain level of injected current (I). In the Shriki model \cite{shriki2003rate}, a strong transient potassium A-current yields a threshold linear f-I curve to a good approximation.

The synaptic weights in all simulations were updated according to the STDP rule presented in \cref{eq:deltaw} with all pre-post spike time pairs contribute additively to the synaptic plasticity.
Further details on the numerical simulations can be found in \cref{sub:HH}.

\begin{figure*}
	\centering 	\hypertarget{foo}{}
		\textbf{STDP dynamics, asymmetric learning rule}\par\medskip
	\includegraphics[width=1\linewidth]{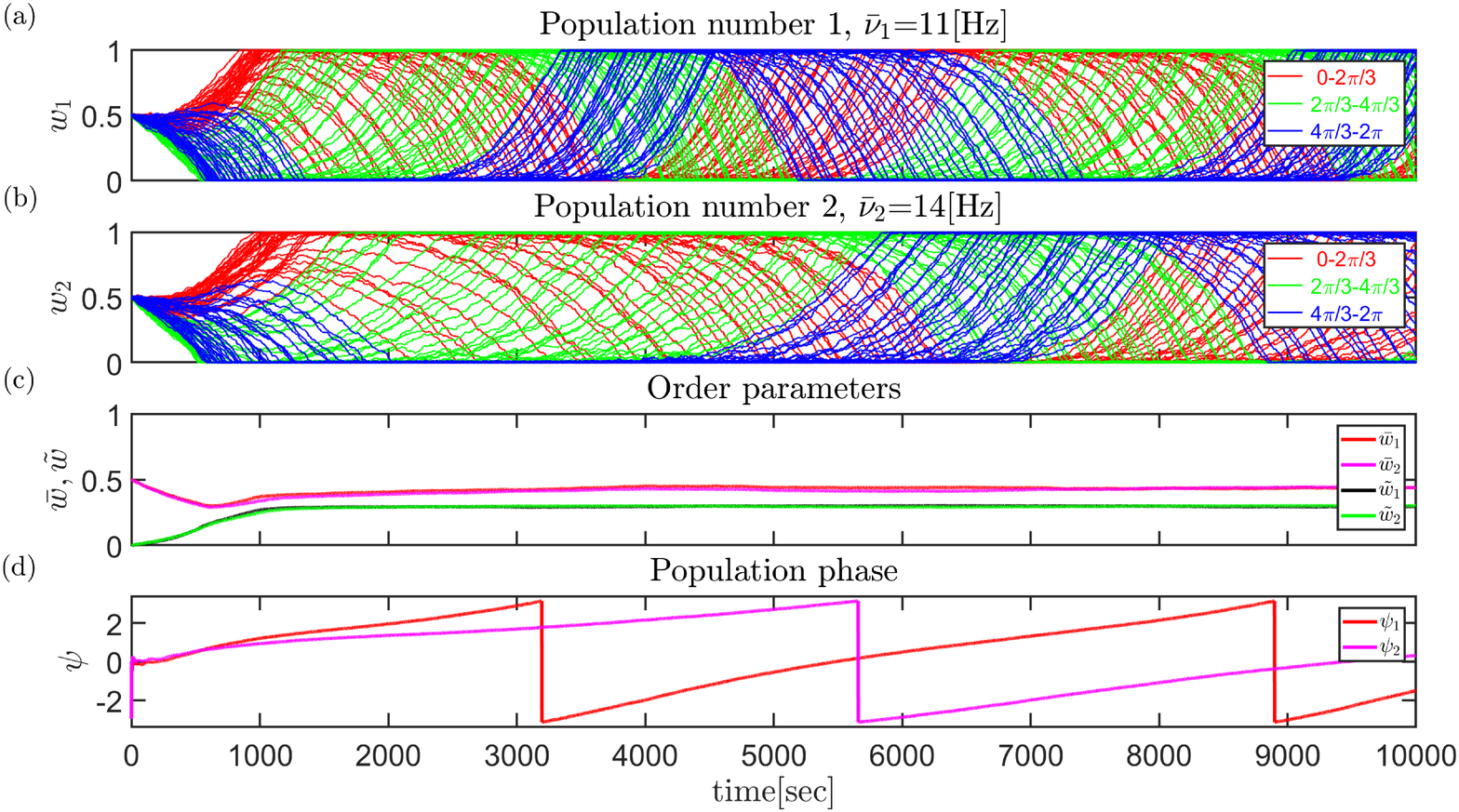}
	\caption{Numerical simulation of STDP dynamics of two populations with 120 excitatory synapses each with the temporally asymmteric STDP rule, \cref{eq:kernel}. (a) and (b) The synaptic weights are shown as a function of time, for population 1 ($11\text{Hz}$) and population 2 ($ 14 \text{Hz}$) in a and b, respectively. The colors represent the preferred phases of the pre-synaptic neurons on the ring. (c) The dynamics of the order parameters, the mean, $\bar{w}$, (red and magenta) and first Fourier component, $\tilde{w}$, (blue and green) are shown as a function of time for both populations. (d) The dynamics of the phases, the phase, $\psi_i$, of each population $i$ is shown as a function of time. Note that the drift velocities are approximately constant in time.}
	\label{fig:wandpsivst}
\end{figure*}

\textbf{Asymmetric learning rule with a strong A-current.}
	\Cref{fig:wandpsivst} presents the results of simulating the STDP dynamics with a conductance based post-synaptic neuron with a strong A-current. In this regime, the f-I curve of the post-synaptic neuron is well approximated by a threshold-linear function (see fig. 1 in \cite{shriki2003rate}). Consequently, it is reasonable to expect that our analytical results will hold, in the limit of a slow learning rate. Indeed, even though the initial conditions of all the synapses are uniform, the uniform solution loses its stability, and a structure that shows phase preference begins to emerge, figs. \hyperlink{foo}{5a} and \hyperlink{foo}{5b}. After about 1000 sec of simulation time, the STDP dynamics of each sub-population converges to an approximately periodic solution. The order parameters $\bar{w}$ and $\tilde{w}$  appear to converge to a fixed point, fig. \hyperlink{foo}{5c}, while the phases $\psi_1$  and $\psi_2$  continue to drift with a relatively fixed velocity, fig. \hyperlink{foo}{5d}.  For the specific choice of parameters used in this simulation, the competitive winner-take-all eigenvalue is stable, see inset \cref{fig:LambdaWTAAsymmetric}, whereas the rhythmic eigenvalue is unstable, see colored stars in \cref{fig:lambdatildeAsymmTau}.

\textbf{Symmetric learning rule with a strong A-current.}

\begin{figure*}
	\centering
	 \textbf{STDP dynamics, symmetric learning rule}\par\medskip
	\includegraphics[width=1\linewidth]{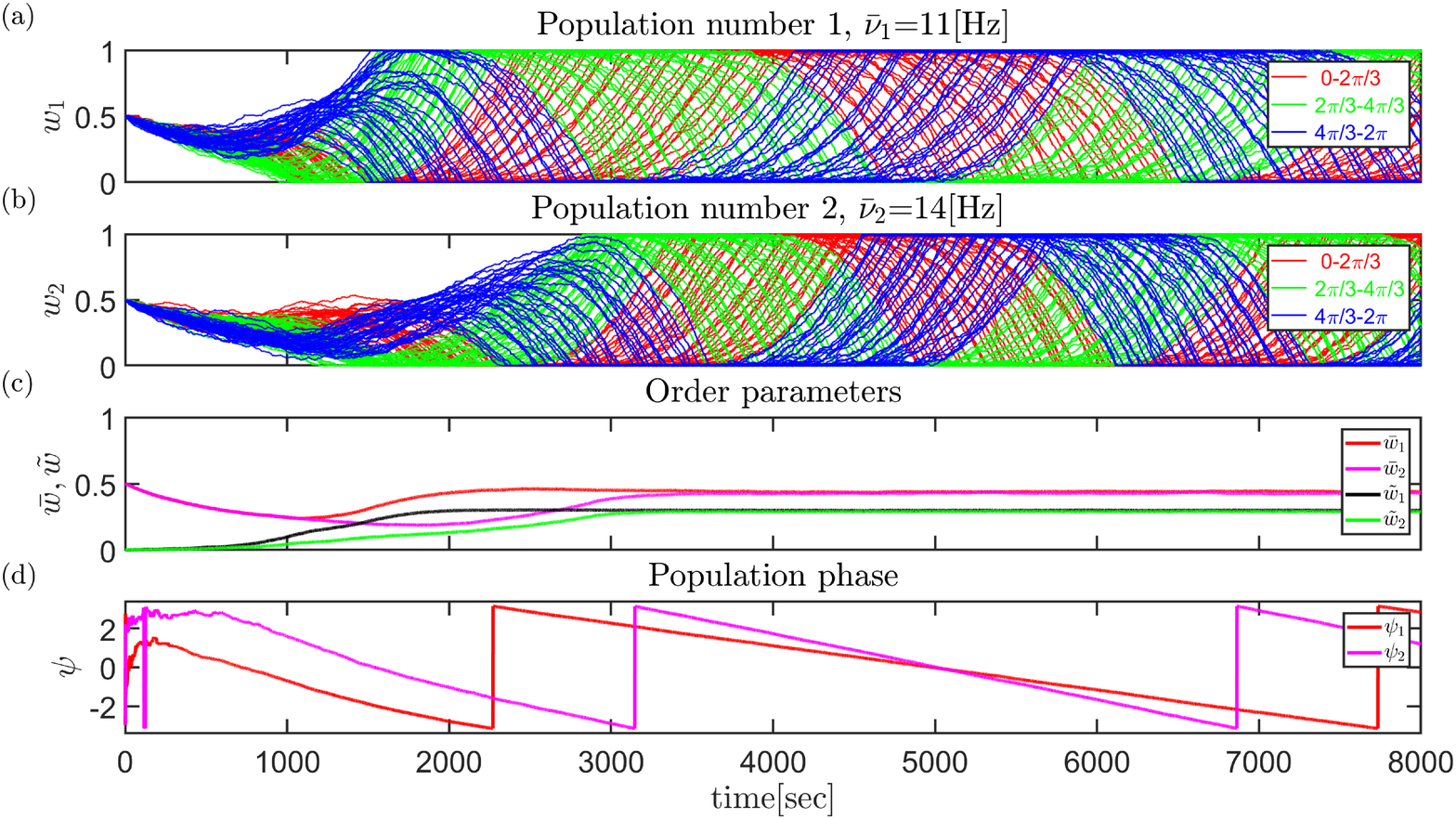}
	\caption{Numerical simulation of STDP dynamics of two populations with 120 excitatory synapses, each  with the  symmetric learning rule, \cref{eq:kernelSymmetric}. (a) and (b) The synaptic weights are shown as a function of time, for population 1 ($11\text{Hz}$) and population 2 ($ 14 \text{Hz}$) in a and b, respectively. (c) The dynamics of the order parameters, the mean, $\bar{w}$, (red and magenta) and first Fourier component, $\tilde{w}$, (blue and green) of each population are shown as a function of time. (d) The dynamics of the phases, the phase, $\psi_i$, of each population $i$ is shown as a function of time.}
	\label{fig:wvstime11and14hzsymmetric}
\end{figure*}

A typical result of simulating the STDP dynamics for the symmetric learning rule is shown in \cref{fig:wvstime11and14hzsymmetric}. As above, the post-synaptic neuron was characterized by a strong A-current, yielding a (threshold) linear f-I curve. In this example, the uniform initial conditions of the synaptic weights evolve to an approximately limit cycle solution for each population after $\sim 1000$ sec. As in the asymmetric case, both populations coexist; hence, STDP facilitates rhythmic information for both learning rules. As before, for the specific choice of parameters used in this simulation the competitive winner-take-all eigenvalue is stable and the rhythmic eigenvalue is unstable, see colored stars in \cref{fig:lambdatildeSymmTau}.

\begin{figure*}
	\centering  	\hypertarget{fii}{}
		\textbf{Synaptic plasticity Vs time, asymmetric learning rule, non-linear neuron}\par\medskip
	\includegraphics[width=1\linewidth]{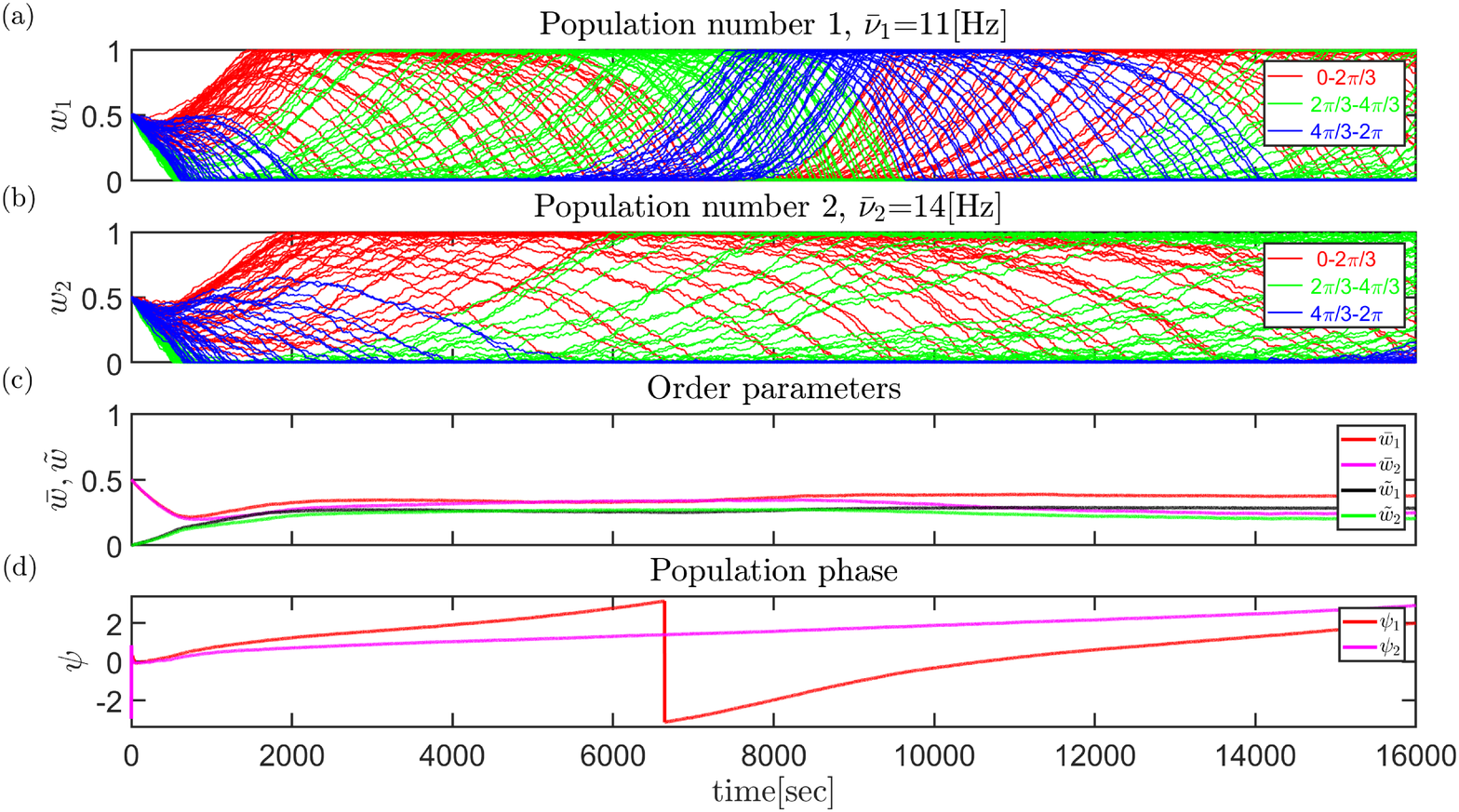}
	\caption{Numerical simulation of the STDP dynamics of two populations with 120 excitatory synapses each  with the  asymmetric learning rule, \cref{eq:kernel} and a non-linear post-synaptic neuron.
	(a) and (b). The synaptic weights are shown as a function of time, for population 1 ($11\text{Hz}$) and population 2 ($ 14 \text{Hz}$) in a and b, respectively. (c) The dynamics of the order parameters- the mean, $\bar{w}$, (red and magenta) and first Fourier component, $\tilde{w}$, (blue and green) of each population are shown as a function of time. (d) The dynamics of the phases: the phase, $\psi_i$, of each population $i$ is shown as a function of time.}
	\label{fig:1114hzasymmetricnonlinear}
\end{figure*}

\textbf{Asymmetric learning rule with a non linear neuron}.
Above we studied the STDP dynamics with a linear post-synaptic neuron numerically. This choice limits the interaction between the two populations. However, it is not uncommon for neurons to have a non-linear f-I curve. In order to determine whether STDP can facilitate multiplexing despite the expected interaction due to the non-linear f-I curve, we used the Shriki model \cite{shriki2003rate} with no A-current. A typical simulation result is shown in \cref{fig:1114hzasymmetricnonlinear}.As can be seen from the figure, the system converges to a dynamical solution that is qualitatively similar to the case of the linear neuron model (\cref{fig:wandpsivst}). Specifically, the order parameters of each synaptic population converge to a solution that enables the transmission of rhythmic activity downstream, whereas the synaptic weights themselves in each population converge to a dynamic solution that appears to be  approximately a limit cycle.

\section{\label{sec:Discussion} Summary \& Discussion}
Rhythmic activity in the brain has fascinated and puzzled neuroscientists for more than a century. Nevertheless, the utility of rhythmic activity remains enigmatic.  One explanation frequently put forward is that of the multiplexing of information. Our work provides some measure of support for this hypothesis from the theory of unsupervised learning. 

We studied the computational implications of a microscopic learning rule, namely STDP, in the absence of a reward or teacher signal. Previous work shows that STDP generates a strong winner-take-all like competition between subgroups of correlated neurons, thus effectively eliminating the possibility of multiplexing \cite{gutig2003learning}. Our work demonstrates that rhythmic activity does not necessarily generate competition between different rhythmic channels. Moreover, we found that under a wide range of parameters STDP dynamics will develop spontaneously the capacity for multiplexing.

Not every learning rule; i.e., a choice of parameters that describes the STDP update rule, will support multiplexing. This observation provides a natural test for our theory. Clearly,  if multiplexing has  evolved via a process of STDP,  then the STDP rule must exhibit instability with respect to the rhythmic modes and stability against fluctuations in the winner-take-all direction. These constraints serve as the basic predictions for our theory.

In Luz \& Shamir \cite{luz2016oscillations}, due to the underlying U(1) symmetry, the system converged to a limit cycle solution. For a single population, the order parameter, $\tilde{w} e^{i \psi}$, will drift on the ring $|\mathbf{w}|=const$  with a constant velocity. Here, in the linear neuron model, in the limit of a slow learning rate one expects the system to converge to the product space of two limit cycles. As there is no reason to expect that the ratio of drift velocities of the two populations will be 	rational, the order parameters $\tilde{w}_1  e^{i \psi_1}$, $\tilde{w}_2  e^{i \psi_2}$  will, most likely, cover the torus uniformly. Nevertheless, the reduced dynamics of each population will exhibit a limit cycle. This intuition relies on the lack of interaction between $\tilde{w}^1$ and $\tilde{w}^2$ in the dynamics of the order parameters, \cref{eq:wdottilde}. Essentially, the interaction between the two populations is mediated solely via their mean component, $\bar{w}$. However, introducing non-linearity to the response properties of the post-synaptic neuron will induce an interaction between the modes. Similarly, for any finite learning rate, $\lambda \neq 0$, the rhythmic modes will not be orthogonal and consequently will be correlated.

Introducing a post-synaptic neuron with a non-linear f-I curve and a finite learning rate induces an interaction between the two populations. Consequently, for a finite learning rate the order parameters  $\tilde{w}_1  e^{i \psi_1}$, $\tilde{w}_2  e^{i \psi_2}$  will not be confined to a torus and  $\tilde{w}_\eta  e^{i \psi_\eta}$  will fluctuate around (in contrast with on) the ring. Traces for this behaviour can be seen by the fact that drift velocity in the numerical simulations is not constant and the global order parameters exhibit small fluctuations around, see \hyperlink{fii}{7c} $\&$  \hyperlink{fii}{7d} for example. Nevertheless, the post-synaptic neuron will respond to both rhythms; hence, multiplexing will be retained.

Synaptic weights in the central nervous system are highly volatile and demonstrate high turnover rates as well as considerable size changes that correlate with the synaptic weight  \cite{mongillo2017intrinsic,attardo2015impermanence,grutzendler2002long,zuo2005development,holtmaat2005transient,holtmaat2006experience,loewenstein2011multiplicative,loewenstein2015predicting}.
How can the brain retain functionality in the face of these considerable changes in synaptic connectivity?  Our work demonstrates how functionality, in terms of retaining the ability to transmit downstream rhythmic information in several channels, can be retained even when the entire synaptic population is modified throughout its entire dynamic range. Here, robustness of function is ensured by stable dynamics for the global order parameters.

\section{\label{sec:Appendix}Appendix}

\subparagraph{The conductance based model for the Post-synaptic neuron}\label{sub:HH}

We used the conductance based model of Shriki et al. \cite{shriki2003rate}.
The model is fully defined and its response to synchronous inputs is discussed in \cite{shriki2003rate}. 

%
Here, we introduce the main features and equations utilized in this model. The voltage equation is 
%
\begin{equation}\label{eq:VdotHHFull}
C_m \dot{V}=-I_{L}-I_{\text{Na}}-I_{k}-I_{A}+\sum_{i}g_i(E_i-V),
\end{equation}

where $I_L$, the leak current is given by $I_L=g_L(V-E_L)$. $I_{Na}$ and $I_k$ are the sodium and potassium currents respectively and are given by $I_{Na}=\bar{g}_{Na}m_{\infty}^3 h(V-E_{Na})$ and $I_{k}=\bar{g}_{k}n^4 h(V-E_{k})$.
The relaxation equations of the the gating variables $x=h,n$ are $dx/dt=(x_\infty-x)/\tau_x$. The time independent functions $x_\infty=h_\infty, n_\infty, m_\infty$ and $\tau_x$ are: $x_\infty=\alpha_x/(\alpha_x+\beta_x)$ and $\tau_x=0.1/(\alpha_x+\beta_x)$, with $\alpha_m=-0.1(V+30)/(exp(-0.1(V+30))-1)$, $\beta_m=4 exp(-(V+55)/18)$,  $\alpha_h=0.07 exp(-(V+44)/20)$ ,  $\beta_h=1/(exp(-0.1(V+14))+1)$, $\alpha_n=-0.01(V+34)/(exp(-0.1(V+34))-1)$ and $\beta_n=0.125 exp(-(V+44)/80)$. 

The A-current  $I_A$ that linearizes the f-I relationship is $I_{A}=\bar{g}_{A}a_{\infty}^3 b(V-E_{k})$, where $a_\infty=1/(exp(-(V+50)/20)+1)$ and $db/dt=(b_\infty-b)/ \tau_A$. The time independent function of $b_\infty$ is $b_\infty=1/(exp((V+80)/6)+1)$ with the  voltage independent time constant $\tau_A$. 
 In the case of the non-linear f-I curve, \cref{fig:1114hzasymmetricnonlinear}, we took this value to be 0.

The term $g_i$ is the total conductance of the $i$th pre-synaptic population and can be written as follows

\begin{equation}\label{cond}
g_x(t)=g_x^0\sum_{j=1}^{N_x}(w_j^x(t)\sum_{s}\frac{[t-t_j^s]_+}{\tau_x} e^{-(t-t_j^s)/\tau_x} ).
\end{equation}
Here, $N_x$ is the number of neurons in the $x$th population, $w_j^x(t)$ is the synaptic weight of the $j$th neuron from the $x$th population and $[y]_+=max(0,y)$. The $s$ spike of the $j$th neuron is denoted by  $t_j^s$.
We used $g_x^0=g_x^R S_x$ with $S_x=1000/N_x$ and $g_x^R=90 \text{mS/cm}^2$ (see  \cite{gutig2003learning,luz2016oscillations} for details).

The membrane capacity is $c_m=0.1 \mu \text{F/cm}^2$. The sodium, potassium and leak conductances are $\bar{g}_{Na}=100 \text{mS/cm}^2$, $\bar{g}_{k}=40  \text{mS/cm}^2$ and  $g_{L}=0.05 \text{mS/cm}^2$  respectively. The conductance and characteristic decay times of the A-current are $g_A=20 \text{mS/cm}^2$ and $\tau_A=20 \text{msec}$ respectively.
The reversal potentials of the ionic and
synaptic currents are $E_{Na} = 55 \text{mV}$, $E_K = −80 \text{mV}$, $E_L = −65 \text{mV}$, $E_e = 0 \text{mV}$, and $E_{in} = −80 \text{mV}$.

\subparagraph{Modeling pre-synaptic activity}

During the spiking neuron simulations, pre-synaptic
activities were modeled by independent inhomogeneous Poisson processes, with a modulated mean firing rate given by \cref{eq:meanfiring}, where $\gamma=1$, and each second $D$ was independently sampled from a uniform distribution with a minimum of $7 \,\text{spikes} \,\times\text{sec}^{-1}$ and a maximum of  $13 \,\text{spikes} \,\times\text{sec}^{-1}$,  ($D=7 + \textbf{U}(0,6) \, \text{spikes} \,\times
\text{sec}^{-1}$). Each pre-synaptic neuron, spiked according to an approximated Bernoulli process, with a probability of $p\approx r \Delta t$, where $r$ is the  mean firing rate (\cref{eq:meanfiring})  and $\Delta t = 1\text{msec}$. The number of pre-synaptic neurons in each population was $N=120$, all excitatory.

\subparagraph{STDP}\label{stdpSim:HH}

The learning rate of the simulations presented in this paper is $\lambda=0.01$. The power of the synaptic weights in \cref{eq:fplusminus} is $\mu=0.01$ and the ratio of depression to potentiation is $\alpha =1.1$. In order to update the synaptic weights
we relied on the separation of time scales between the synaptic dynamics of \cref{eq:VdotHHFull}; hence, the synaptic weights were updated every $1 \text{sec}$ of simulation.

\begin{itemize}
	
	\item{\textbf{Asymmetric learning rule}} The characteristic decay times were chosen to be $\tau_{+}=20 \text{ms}$ and $\tau_{-}=50 \text{ms}$.

		\item{\textbf{Symmetric learning rule}} Here,  based on our analysis, we chose the ratio of decay times to be $\sim 10$, $\tau_{+}=5 \text{ms}$, $\tau_{-}=50 \text{ms}$.   

\end{itemize}

In addition, the initial conditions for all neurons were uniform; i.e., $w_\eta(\phi,t=0)=0.5$, $\eta=1,2$.

\section*{\label{sec:ACKNOWLEDGMENTS}ACKNOWLEDGMENTS}
This research was supported by the Israel Science Foundation (ISF) grant number 300/16, and in part by the National Science Foundation under Grant No. NSF PHY-1748958 and the United States-Israel Binational Science Foundation grant 2013204. \newpage

\bibliographystyle{unsrt}
\bibliography{Multiplexing_rhythmic_information_by_spike_timing_dependent_plasticity}

\begin{thebibliography}{10}

\bibitem{coenen2014adolf}
Anton Coenen, Edward Fine, and Oksana Zayachkivska.
\newblock Adolf beck: A forgotten pioneer in electroencephalography.
\newblock {\em Journal of the History of the Neurosciences}, 23(3):276--286,
  2014.

\bibitem{Haas2003}
L.~Haas.
\newblock Hans berger (1873-1941), richard caton (1842-1926), and
  electroencephalography.
\newblock {\em J Neurol Neurosurg Psychiatry}, 74(1):9--9, Jan 2003.
\newblock 12486257[pmid].

\bibitem{steriade1985abolition}
M~Steriade, M~Deschenes, L~Domich, and C~Mulle.
\newblock Abolition of spindle oscillations in thalamic neurons disconnected
  from nucleus reticularis thalami.
\newblock {\em Journal of neurophysiology}, 54(6):1473--1497, 1985.

\bibitem{buzsaki1992high}
Gyorgy Buzsaki, Zsolt Horvath, Ronald Urioste, Jamille Hetke, and Kensall Wise.
\newblock High-frequency network oscillation in the hippocampus.
\newblock {\em Science}, 256(5059):1025--1027, 1992.

\bibitem{gray1994synchronous}
Charles~M Gray.
\newblock Synchronous oscillations in neuronal systems: mechanisms and
  functions.
\newblock {\em Journal of computational neuroscience}, 1(1-2):11--38, 1994.

\bibitem{bragin1999high}
Anatol Bragin, Jerome Engel~Jr, Charles~L Wilson, Itzhak Fried, and Gyorgy
  Buzs{\'a}ki.
\newblock High-frequency oscillations in human brain.
\newblock {\em Hippocampus}, 9(2):137--142, 1999.

\bibitem{burgess2002functional}
Adrian~P Burgess and Lia Ali.
\newblock Functional connectivity of gamma eeg activity is modulated at low
  frequency during conscious recollection.
\newblock {\em International Journal of Psychophysiology}, 46(2):91--100, 2002.

\bibitem{buzsaki2006rhythms}
Gyorgy Buzsaki.
\newblock {\em Rhythms of the Brain}.
\newblock Oxford University Press, 2006.

\bibitem{shamir2009representation}
Maoz Shamir, Oded Ghitza, Steven Epstein, and Nancy Kopell.
\newblock Representation of time-varying stimuli by a network exhibiting
  oscillations on a faster time scale.
\newblock {\em PLoS computational biology}, 5(5):e1000370, 2009.

\bibitem{buzsaki2015editorial}
Gy{\"o}rgy Buzs{\'a}ki and Walter Freeman.
\newblock Editorial overview: brain rhythms and dynamic coordination.
\newblock {\em Current opinion in neurobiology}, 31:v--ix, 2015.

\bibitem{ray2015gamma}
Supratim Ray and John~HR Maunsell.
\newblock Do gamma oscillations play a role in cerebral cortex?
\newblock {\em Trends in cognitive sciences}, 19(2):78--85, 2015.

\bibitem{bocchio2017synaptic}
Marco Bocchio, Sadegh Nabavi, and Marco Capogna.
\newblock Synaptic plasticity, engrams, and network oscillations in amygdala
  circuits for storage and retrieval of emotional memories.
\newblock {\em Neuron}, 94(4):731--743, 2017.

\bibitem{Proskovec2018}
Amy~L. Proskovec, Alex~I. Wiesman, Elizabeth Heinrichs-Graham, and Tony~W.
  Wilson.
\newblock Beta oscillatory dynamics in the prefrontal and superior temporal
  cortices predict spatial working memory performance.
\newblock {\em Scientific Reports}, 8(1):8488, 2018.

\bibitem{taub2018oscillations}
Aryeh~Hai Taub, Rita Perets, Eilat Kahana, and Rony Paz.
\newblock Oscillations synchronize amygdala-to-prefrontal primate circuits
  during aversive learning.
\newblock {\em Neuron}, 97(2):291--298, 2018.

\bibitem{engel1992temporal}
Andreas~K Engel, Peter K{\"o}nig, Andreas~K Kreiter, Thomas~B Schillen, and
  Wolf Singer.
\newblock Temporal coding in the visual cortex: new vistas on integration in
  the nervous system.
\newblock {\em Trends in neurosciences}, 15(6):218--226, 1992.

\bibitem{singer1995visual}
Wolf Singer and Charles~M Gray.
\newblock Visual feature integration and the temporal correlation hypothesis.
\newblock {\em Annual review of neuroscience}, 18(1):555--586, 1995.

\bibitem{engel2001dynamic}
Andreas~K Engel, Pascal Fries, and Wolf Singer.
\newblock Dynamic predictions: oscillations and synchrony in top--down
  processing.
\newblock {\em Nature Reviews Neuroscience}, 2(10):704, 2001.

\bibitem{fries2005mechanism}
Pascal Fries.
\newblock A mechanism for cognitive dynamics: neuronal communication through
  neuronal coherence.
\newblock {\em Trends in cognitive sciences}, 9(10):474--480, 2005.

\bibitem{jensen2007human}
Ole Jensen, Jochen Kaiser, and Jean-Philippe Lachaux.
\newblock Human gamma-frequency oscillations associated with attention and
  memory.
\newblock {\em Trends in neurosciences}, 30(7):317--324, 2007.

\bibitem{knyazev2007motivation}
Gennady~G Knyazev.
\newblock Motivation, emotion, and their inhibitory control mirrored in brain
  oscillations.
\newblock {\em Neuroscience \& Biobehavioral Reviews}, 31(3):377--395, 2007.

\bibitem{hobson2002cognitive}
J~Allan Hobson and Edward~F Pace-Schott.
\newblock The cognitive neuroscience of sleep: neuronal systems, consciousness
  and learning.
\newblock {\em Nature Reviews Neuroscience}, 3(9):679, 2002.

\bibitem{engel2010beta}
Andreas~K Engel and Pascal Fries.
\newblock Beta-band oscillations—signalling the status quo?
\newblock {\em Current opinion in neurobiology}, 20(2):156--165, 2010.

\bibitem{storchi2017modulation}
Riccardo Storchi, Robert~A Bedford, Franck~P Martial, Annette~E Allen, Jonathan
  Wynne, Marcelo~A Montemurro, Rasmus~S Petersen, and Robert~J Lucas.
\newblock Modulation of fast narrowband oscillations in the mouse retina and
  dlgn according to background light intensity.
\newblock {\em Neuron}, 93(2):299--307, 2017.

\bibitem{Hebb}
Donald~Olding Hebb.
\newblock {\em The organization of behavior: A neuropsychological theory}.
\newblock Psychology Press, 2005.

\bibitem{luz2016oscillations}
Yotam Luz and Maoz Shamir.
\newblock Oscillations via spike-timing dependent plasticity in a feed-forward
  model.
\newblock {\em PLoS computational biology}, 12(4):e1004878, 2016.

\bibitem{CFCExample1}
Takumi Sase, Yuichi Katori, Motomasa Komuro, and Kazuyuki Aihara.
\newblock Bifurcation analysis on phase-amplitude cross-frequency coupling in
  neural networks with dynamic synapses.
\newblock {\em Frontiers in Computational Neuroscience}, 11:18, 2017.

\bibitem{CFCExample2}
Alexandre Hyafil, Anne-Lise Giraud, Lorenzo Fontolan, and Boris Gutkin.
\newblock Neural cross-frequency coupling: connecting architectures,
  mechanisms, and functions.
\newblock {\em Trends in neurosciences}, 38(11):725--740, 2015.

\bibitem{CFCExample3}
Anita~K Roopun, Mark~A Kramer, Lucy~M Carracedo, Marcus Kaiser, Ceri~H Davies,
  Roger~D Traub, Nancy~J Kopell, and Miles~A Whittington.
\newblock Temporal interactions between cortical rhythms.
\newblock {\em Frontiers in neuroscience}, 2:34, 2008.

\bibitem{CFCExample4}
Felix Siebenh{\"u}hner, Sheng~H Wang, J~Matias Palva, and Satu Palva.
\newblock Cross-frequency synchronization connects networks of fast and slow
  oscillations during visual working memory maintenance.
\newblock {\em Elife}, 5:e13451, 2016.

\bibitem{saleem2017subcortical}
Aman~B Saleem, Anthony~D Lien, Michael Krumin, Bilal Haider, Miroslav~Roman
  Roson, Asli Ayaz, Kimberly Reinhold, Laura Busse, Matteo Carandini, and
  Kenneth~D Harris.
\newblock Subcortical source and modulation of the narrowband gamma oscillation
  in mouse visual cortex.
\newblock {\em Neuron}, 93(2):315--322, 2017.

\bibitem{gilson2009emergence}
Matthieu Gilson, Anthony~N Burkitt, David~B Grayden, Doreen~A Thomas, and J~Leo
  van Hemmen.
\newblock Emergence of network structure due to spike-timing-dependent
  plasticity in recurrent neuronal networks iii: Partially connected neurons
  driven by spontaneous activity.
\newblock {\em Biological Cybernetics}, 101(5-6):411, 2009.

\bibitem{gutig2003learning}
Robert G{\"u}tig, Ranit Aharonov, Stefan Rotter, and Haim Sompolinsky.
\newblock Learning input correlations through nonlinear temporally asymmetric
  hebbian plasticity.
\newblock {\em Journal of Neuroscience}, 23(9):3697--3714, 2003.

\bibitem{morrison2008phenomenological}
Abigail Morrison, Markus Diesmann, and Wulfram Gerstner.
\newblock Phenomenological models of synaptic plasticity based on spike timing.
\newblock {\em Biological cybernetics}, 98(6):459--478, 2008.

\bibitem{song2000competitive}
Sen Song, Kenneth~D Miller, and Larry~F Abbott.
\newblock Competitive hebbian learning through spike-timing-dependent synaptic
  plasticity.
\newblock {\em Nature neuroscience}, 3(9):919, 2000.

\bibitem{shamir2019theories}
Maoz Shamir.
\newblock Theories of rhythmogenesis.
\newblock {\em Current opinion in neurobiology}, 58:70--77, 2019.

\bibitem{luz2014effect}
Yotam Luz and Maoz Shamir.
\newblock The effect of stdp temporal kernel structure on the learning dynamics
  of single excitatory and inhibitory synapses.
\newblock {\em PloS one}, 9(7):e101109, 2014.

\bibitem{morrison2008}
Abigail Morrison, Markus Diesmann, and Wulfram Gerstner.
\newblock Phenomenological models of synaptic plasticity based on spike timing.
\newblock {\em Biological cybernetics}, 98(6):459--478, 2008.

\bibitem{kempter2001intrinsic}
Richard Kempter, Wulfram Gerstner, and J~Leo~van Hemmen.
\newblock Intrinsic stabilization of output rates by spike-based hebbian
  learning.
\newblock {\em Neural computation}, 13(12):2709--2741, 2001.

\bibitem{kempter1999hebbian}
Richard Kempter, Wulfram Gerstner, and J~Leo Van~Hemmen.
\newblock Hebbian learning and spiking neurons.
\newblock {\em Physical Review E}, 59(4):4498, 1999.

\bibitem{luz2012balancing}
Yotam Luz and Maoz Shamir.
\newblock Balancing feed-forward excitation and inhibition via hebbian
  inhibitory synaptic plasticity.
\newblock {\em PLoS computational biology}, 8(1):e1002334, 2012.

\bibitem{markram1997regulation}
Henry Markram, Joachim L{\"u}bke, Michael Frotscher, and Bert Sakmann.
\newblock Regulation of synaptic efficacy by coincidence of postsynaptic aps
  and epsps.
\newblock {\em Science}, 275(5297):213--215, 1997.

\bibitem{bi1998synaptic}
Guo-qiang Bi and Mu-ming Poo.
\newblock Synaptic modifications in cultured hippocampal neurons: dependence on
  spike timing, synaptic strength, and postsynaptic cell type.
\newblock {\em Journal of neuroscience}, 18(24):10464--10472, 1998.

\bibitem{sjostrom2001rate}
Per~Jesper Sj{\"o}str{\"o}m, Gina~G Turrigiano, and Sacha~B Nelson.
\newblock Rate, timing, and cooperativity jointly determine cortical synaptic
  plasticity.
\newblock {\em Neuron}, 32(6):1149--1164, 2001.

\bibitem{zhang1998critical}
Li~I Zhang, Huizhong~W Tao, Christine~E Holt, William~A Harris, and Mu-ming
  Poo.
\newblock A critical window for cooperation and competition among developing
  retinotectal synapses.
\newblock {\em Nature}, 395(6697):37, 1998.

\bibitem{Abbott2000}
L.~F. Abbott and Sacha~B. Nelson.
\newblock Synaptic plasticity: taming the beast.
\newblock {\em Nature Neuroscience}, 3(11):1178--1183, 2000.

\bibitem{froemke2006contribution}
Robert~C Froemke, Ishan~A Tsay, Mohamad Raad, John~D Long, and Yang Dan.
\newblock Contribution of individual spikes in burst-induced long-term synaptic
  modification.
\newblock {\em Journal of neurophysiology}, 95(3):1620--1629, 2006.

\bibitem{Nishiyama2000CalciumSR}
Makoto Nishiyama, Kyonsoo Hong, Katsuhiko Mikoshiba, Mu-Ming Poo, and Kunio
  Kato.
\newblock Calcium stores regulate the polarity and input specificity of
  synaptic modification.
\newblock {\em Nature}, 408:584--588, 2000.

\bibitem{shouval2002unified}
Harel~Z Shouval, Mark~F Bear, and Leon~N Cooper.
\newblock A unified model of nmda receptor-dependent bidirectional synaptic
  plasticity.
\newblock {\em Proceedings of the National Academy of Sciences},
  99(16):10831--10836, 2002.

\bibitem{WOODIN2003807}
Melanie~A Woodin, Karunesh Ganguly, and Mu~ming Poo.
\newblock Coincident pre- and postsynaptic activity modifies gabaergic synapses
  by postsynaptic changes in cl− transporter activity.
\newblock {\em Neuron}, 39(5):807 -- 820, 2003.

\bibitem{shriki2003rate}
Oren Shriki, David Hansel, and Haim Sompolinsky.
\newblock Rate models for conductance-based cortical neuronal networks.
\newblock {\em Neural computation}, 15(8):1809--1841, 2003.

\bibitem{mongillo2017intrinsic}
Gianluigi Mongillo, Simon Rumpel, and Yonatan Loewenstein.
\newblock Intrinsic volatility of synaptic connections—a challenge to the
  synaptic trace theory of memory.
\newblock {\em Current opinion in neurobiology}, 46:7--13, 2017.

\bibitem{attardo2015impermanence}
Alessio Attardo, James~E Fitzgerald, and Mark~J Schnitzer.
\newblock Impermanence of dendritic spines in live adult ca1 hippocampus.
\newblock {\em Nature}, 523(7562):592, 2015.

\bibitem{grutzendler2002long}
Jaime Grutzendler, Narayanan Kasthuri, and Wen-Biao Gan.
\newblock Long-term dendritic spine stability in the adult cortex.
\newblock {\em Nature}, 420(6917):812, 2002.

\bibitem{zuo2005development}
Yi~Zuo, Aerie Lin, Paul Chang, and Wen-Biao Gan.
\newblock Development of long-term dendritic spine stability in diverse regions
  of cerebral cortex.
\newblock {\em Neuron}, 46(2):181--189, 2005.

\bibitem{holtmaat2005transient}
Anthony~JGD Holtmaat, Joshua~T Trachtenberg, Linda Wilbrecht, Gordon~M
  Shepherd, Xiaoqun Zhang, Graham~W Knott, and Karel Svoboda.
\newblock Transient and persistent dendritic spines in the neocortex in vivo.
\newblock {\em Neuron}, 45(2):279--291, 2005.

\bibitem{holtmaat2006experience}
Anthony Holtmaat, Linda Wilbrecht, Graham~W Knott, Egbert Welker, and Karel
  Svoboda.
\newblock Experience-dependent and cell-type-specific spine growth in the
  neocortex.
\newblock {\em Nature}, 441(7096):979, 2006.

\bibitem{loewenstein2011multiplicative}
Yonatan Loewenstein, Annerose Kuras, and Simon Rumpel.
\newblock Multiplicative dynamics underlie the emergence of the log-normal
  distribution of spine sizes in the neocortex in vivo.
\newblock {\em Journal of Neuroscience}, 31(26):9481--9488, 2011.

\bibitem{loewenstein2015predicting}
Yonatan Loewenstein, Uri Yanover, and Simon Rumpel.
\newblock Predicting the dynamics of network connectivity in the neocortex.
\newblock {\em Journal of Neuroscience}, 35(36):12535--12544, 2015.

\end{thebibliography}
\end{document}